\documentclass[fleqn,usenatbib]{mnras}

\usepackage[T1]{fontenc}

\DeclareRobustCommand{\VAN}[3]{#2}
\let\VANthebibliography\thebibliography
\def\thebibliography{\DeclareRobustCommand{\VAN}[3]{##3}\VANthebibliography}

\usepackage{graphicx}	
\usepackage{amsmath}	
\usepackage{subcaption}
\usepackage{booktabs}

\def\vmicro{$\xi_{\rm t}$}
\def\vmacro{$\zeta_{\rm RT}$}

\newcommand{\teff}{$T_{\rm eff}$}
\newcommand{\kms}{km\,s$^{-1}$}
\newcommand{\lgg}{$\log{g}$}
\newcommand{\vsini}{$v\sin{i}$}

\newcommand{\vs}{$v_{\rm e}\sin i$}
\newcommand{\abun}{$\log\varepsilon$}
\def\bs{$\langle B_{\rm s} \rangle$}

\newcommand{\SV}{\textsc{SynthVb}}
\newcommand{\bmag}{\textsc{BinMag6}}

\newcommand{\SME}{\textsc{SME}}
\newcommand{\SED}{\textsc{SED}}

\def\ione{\,{\sc i}}
\def\ii{\,{\sc ii}}
\def\iii{\,{\sc iii}}
\newcommand\omicron{o}
\def\Gem{$\gamma$~Gem}
\def\Peg{$\omicron$~Peg}
\def\Vir{$\theta$~Vir}
\def\Cap{$\nu$~Cap}

\title[Do A-B stars with normal abundances exist?]{Non-LTE abundance analysis of A-B stars with low rotational velocities. II. Do A-B stars with normal abundances exist?}

\author[A. M. Romanovskaya et al.]
{
\parbox{\textwidth}{
A. M. Romanovskaya$^{1}$,\thanks{E-mail: annarom@inasan.ru}
T. A. Ryabchikova$^{1}$,
Yu. V. Pakhomov$^{1}$,
S. A. Korotin$^{2}$,
T. M. Sitnova$^{1}$}
\\
\\
$^{1}$Institute of Astronomy, Russian Academy of Sciences, Pyatnitskaya 48, 119017, Moscow, Russia\\
$^{2}$Physics of stars department, Crimean Astrophysical Observatory,
Nauchny 298409, Republic of Crimea\\ 
}

\date{Accepted XXX. Received YYY; in original form ZZZ}

\pubyear{2023}

\begin{document}
\label{firstpage}
\pagerange{\pageref{firstpage}--\pageref{lastpage}}
\maketitle


\begin{abstract}
We present chemical composition and fundamental parameters (the effective temperature, surface gravity and radius) for four sharp-lined A-type stars \Gem\ (HD~41705), \Peg\ (HD~214994), \Vir\ (HD~114330) and \Cap\ (HD~193432). Our analysis is based on a self-consistent model fitting of high-resolution spectra and spectrophotometric observations over a wide wavelength range. 
We refined the fundamental parameters of the stars with the \SME\, package and verified their accuracy by comparing with the spectral energy distribution and hydrogen line profiles. We found  \teff/\lgg\, = 9190$\pm$130~K/3.56$\pm$0.08, 9600$\pm$50~K/3.81$\pm$0.04, 9600$\pm$140~K/3.61$\pm$0.12, and 10200$\pm$220~K/3.88$\pm$0.08 for \Gem, \Peg, \Vir\, and \Cap, respectively. Our detailed abundance analysis employs a hybrid technique for spectrum synthesis based on classical model atmospheres calculated in local thermodynamic equilibrium (LTE) assumption together with the non-LTE (NLTE) line formation for 18 of 26 investigated species. 
Comparison of the abundance patterns observed in A stars of different types (normal A, Am, Ap) with similar fundamental parameters reveals significant abundance diversity that cannot be explained by the current mechanisms of abundance peculiarity formation in stellar atmospheres.
We found a rise of the heavy element (Zn, Sr, Y, Zr, Ba) abundance excess up to +1 dex with \teff\ increasing from 7200 to 10000~K, with a further decrease down to solar value at \teff\ = 13000~K, indicating that stars with solar element abundances can be found among late B-type stars.

\end{abstract}




\section{Introduction}\label{Intro}

The chemically peculiar stars of spectral types from A to middle B are known for various types of chemical anomalies. 
They include both chemically-peculiar magnetic (Ap SiSrCrEu type) and non-magnetic (Hg-Mn type) stars as well as stars with moderately enhanced metalic lines (Am) and weakened metalic lines ($\lambda$ Boo type).

The chemical anomalies in peculiar stars are of great interest for studying and understanding the physics of processes in the interiors of stars. 
It is believed that the origin of the chemical anomalies is connected with the atomic diffusion processes. 
To explain the observed anomalies, \citet{1970ApJ...160..641M} proposed a theory of diffusion, according to which the diffusion of atoms and ions of an element occurs under the action of gravity directed towards the centre of the star and radiation pressure forces pushing particles into the outer layers of the atmosphere. 

To explain the Am star phenomenon two different atomic diffusion scenarios are proposed.
\citet{1971A&A....13..263W} suggested that the separation of chemical elements occurs directly below the hydrogen convection zone, in the radiative zone \citep[see e.g.][]{1996A&A...310..872A} and it is assumed that a very small mass of a star gets high abundance when the star is on the main sequence (MS) stage. Further, when the star reaches the subgiant branch, a decrease in abundances in the star's atmosphere is assumed. In another model proposed by \citet{2000ApJ...529..338R}, the separation of elements occurs much deeper in the star. Consequently, larger part of stellar mass gets an anomalous chemical composition when the star is in the MS stage and, further, the anomalies remain presumably longer as the star evolves to the subgiant stage.

To understand the mechanisms of formation of chemical element anomalies, accurate elemental abundance determinations from light CNO to the rare earth elements (lanthanides) are required for each group of chemically peculiar stars.
Observations \citep{1974ARA&A..12..257P} and theory \citep{1982ApJ...258..349M} show that slow rotation (equatorial velocity below 120~\kms) favours the presence of chemical anomalies in hot A-B stars. \citet[][hereafter, Paper~\ione]{2020MNRAS.499.3706M} carried out a detailed analysis of the atmospheres of a sample of slowly rotating A-B stars with effective temperatures mostly above 9300~K.  Abundances of 14 chemical elements from He to Nd were determined taking into account deviations from the local thermodynamic equilibrium (NLTE approach). Both Am and normal A stars have an excess of heavy elements Sr-Zr-Ba-Nd compared to the solar abundance, with Ba having the maximum excess among them. An increase in the abundance of heavier elements seems to correlate with an increase in the effective temperature in hot stars. 

In this study we aim to complement the stellar sample of \citet{2020MNRAS.499.3706M} with effective temperature range from 9000 to 11000~K with a detailed chemical element abundances determined in LTE and NLTE, where available.
This sample includes four A-B stars: \Gem\ (HD~47105, Alhena), \Peg\ (HD~214994), \Vir\ (HD~114330), and \Cap\ (HD~193432). 
Abundances of 26 elements were obtained with the NLTE treatment for 18 of them. 

The paper is organised as follows. Section~\ref{obs} provides information about sample stars observations used in our study.  The methods and the results of the atmospheric parameter determinations are given in Section~\ref{model}. In Section~\ref{abund} we provide the results of abundance analysis. Comparison of our abundance results with those from the literature for the sample stars is given in Section~\ref{Comp1}. Evolutionary status of the sample stars is considered in Section~\ref{evol}. We compare abundance patterns of different types of A stars (normal A, Am, Ap) with close atmospheric parameters and evolutionary status in Section~\ref{Comp2}.  Discussion and conclusions are presented in Section~\ref{conclusions}.

\section{Stellar sample and observations}
\label{obs}

\subsection{Stellar sample}
\label{sec1:star} 
\Gem\, is a well known spectroscopic binary star, where the A component Alhena A has an extremely small surface magnetic field of 30~G  \citep{2020MNRAS.492.5794B}. The mass of the primary is 2.8~$M_{\odot}$. The secondary component is 5 - 6 mag fainter  than the primary star and it has a mass of 1.07 $M_{\odot}$ \citep{1993AJ....106.1156F}. It was classified as a cool G star \citep{2014A&A...572A..91T}. The primary was classified as a normal A type star, but with higher abundances of Zr, Ba, La, Ce, and Nd compared to solar values, and it lies somewhere between a normal A star and Am star \citep{2015PASP..127..340A}. 
\textsc{SIMBAD} database\footnote{http://simbad.u-strasbg.fr/simbad/} contains data on fundamental parameters obtained through different methods. The effective temperature (\teff ) ranges from 8900 K to 10700~K, while the surface gravity (\lgg ) spans from 3.46 to 4.00.

\Peg\, was identified as a low-amplitude, long-period binary with the spectral type of the primary A1 IV. The mass function is $f(m)$ = 0.00066 $\pm$ 0.00024 $M_{\odot}$ \citep{1999ASPC..185..378F}. There is no information about mass and spectral type of the secondary component, but we expect its mass to be much smaller than that of the primary star. \textsc{SIMBAD} provides the effective temperature in the range of 8300 K to 10080 K, and the surface gravity is given within the range of 3.20 to 4.00. No significant polarisation signal was detected in course of the highly sensitive search for magnetic fields in B, A and F stars \citep{2002AA...392..637S}, providing an estimate -32$\pm$20~G for the longitudinal magnetic field.  

\Vir\, was identified as A1~IV spectral type double or multiple star. According to Washington Double Star (\textsc{WDS}) catalog \citep{2001AJ....122.3466M}. $\theta$\,Vir is a multiple system containing four components. Two of them are displayed at a distance of $7''$ and $71''$ from the main component. The main star is an interferometric binary with a separation of $\sim 0.4''$ \citep{1982ApJS...48..273M} and magnitude difference of $2.2^m$ \citep{2000AJ....119.2403T}. Mass estimates of 2.98 and 0.08~$M_{\odot}$ are given in \citet{2009MNRAS.398.2085S} for both components, respectively. The contribution of the secondary component to the spectrum may be not negligible, however, the star was included in the list of CALSPEC spectrophotometric standards \citep{2022ApJS..263....1R}. 
Finally, we ignored the possible effect of a close secondary component on the common spectra of sample stars and on the spectral energy distributions in our analysis.

\Cap\, is a superficially normal single star with B9~IV spectral type. The elemental abundances of \Cap\, was recently analysed in \citep{2018ApJ...854...50M} and \citep{1991MNRAS.252..116A}, but non-LTE effects were neglected. The effective temperature is given in \textsc{SIMBAD} in the range of 9690--10900~K, and the surface gravity is given as 3.00--4.00.

Part of the literature data on the fundamental parameters of the program stars is given in Table~\ref{literat-param}. It is important to note that the majority of the previous abundance studies were conducted in LTE. Here and throughout the paper, an error in the measurement of the last digits is given in parentheses.

\begin{table}                                                                    
\caption{Literature data on the fundamental parameters of the program stars.}       
\small                                                                                
\setlength\tabcolsep{0.8pt}                                                             
\begin{tabular}{lllll}                                                                
\hline \hline Star  & \teff & \lgg & Reference & Method \\                            
\hline                                                                                
\Gem   &~8953		    &	3.46	    &	\citet{2003AJ....126.2048G}&	Spectrophotometry \\     
       &~9040(280) &		        &	\citet{2009AA...501..297Z} &	Spectrophotometry \\     
       &~9150(100) & 3.60(10) & \citet{2002AA...392.1031A} & Spectroscopy \\          
       &~9341      &          & \citet{1998AA...339..858D} & Photometry \\            
       &~9440		    &	3.59	    &	\citet{1993AA...276..142H}	&	Spectroscopy \\          
\hline                                                                                
\Peg   &~9373(303)	&	3.73(23) &	\citet{2011AA...531A.165P}	&	Spectroscopy \\          
       &~9575(15)  & 3.73(01) & \citet{2015PASP..127..340A}& Spectroscopy \\          
       &~9680		    &	3.71	    &	\citet{1993AA...276..142H}	&	Spectroscopy \\          
       &~9720		    &		        &	\citet{1998AA...339..858D}	&	Photometry \\            
       &~9930(350) &		        &	\citet{2009AA...501..297Z}	&	Spectrophotometry \\     
\hline                                                                                
\Vir   &~9500	     &	3.60	    &	\citet{1987Afz....26...55D}&	Spectroscopy \\          
       &~9509		    &	3.80	    &	\citet{2007MNRAS.374..664C}&	Spectroscopy\\           
  	    &~9570(264)	&	3.95(11) &	\citet{2011AA...531A.165P}	&	Spectroscopy \\          
       &~9671(289) &	3.57(82) &	\citet{2012AA...538A.143K}	&	Spectroscopy \\          
\hline                                                                                
\Cap   &10250     	&	3.90     &	\citet{1991MNRAS.252..116A}&	Spectrophotometry \\ 
       &10185	     &	3.88	    &	\citet{2003AA...398.1121E} &	Photometry \\          
       &10300(250)	&	3.90(25)	&	\citet{2018ApJ...854...50M}&	Photometry\\           
\hline                                                                                
\end{tabular}                                                                         
\label{literat-param}                                                                 
\end{table}

\subsection{Observations}
\label{sec1:obs}
For the sample stars, we used spectra taken with different instruments, selecting the highest quality spectrum available in the archives for each star. For \Gem, spectroscopic observations were extracted from the FEROS spectrograph (MPG/ESO 2.2-metre telescope, La Silla) archive\footnote{http://archive.eso.org/scienceportal/home} and cover the spectral range of 3500-9200 \AA\, (program ID -- 082.A-9007, PI -- J. Carson, SNR$\simeq$430). The resolving power of the instrument is $R = \lambda/\delta\lambda = 48\,000$. For \Peg, we used spectra from ESPaDOnS spectrograph (Canada-France-Hawaii Telescope) archive\footnote{https://www.cadc-ccda.hia-iha.nrc-cnrc.gc.ca/en/cfht/} covering the spectral range 3700-10500 \AA\, with $R = 68\,000$ (proposal ID -- 14AC03, PI --  V. Khalack). Twelve spectra were averaged providing the final SNR$\approx$700. For \Vir, we found available observations taken with the HARPS spectrograph (High Accuracy Radial velocity Planet Searcher) archive\footnote{http://archive.eso.org/wdb/wdb/eso/repro/form} (3000-10000 \AA, $R = 80\,000$, program ID -- 60.A-9709(G), SNR$\simeq$250) and we also used ESPaDOnS spectrum for the near infrared (IR) range (proposal ID -- E02, PI -- N. Manset). For \Cap, spectral observations taken with the UVES spectrograph were extracted from the ESO archive (programs ID -- 67.D-0384(A), PI --  J. Orosz; 076.D-0169(A), PI --  Ch. Cowley). Spectra cover 3730-6835 \AA\ wavelength range with $R = 37\,000$, SNR$\simeq$150, and 5650-9460~\AA\ wavelength range with $R = 107\,000$ and SNR$\simeq$250. The last spectrum was used for nitrogen and oxygen abundance determination.

Photometric and spectrophotometric observations in different spectral regions were employed to construct the spectral energy distribution (SED) for each star. For the ultraviolet (UV) region, we used observations obtained with the S2/68 telescope of TD1 mission (European Space Research Organization (ESRO) satellite), which measured the absolute flux in four bands, 1565 \AA, 1965 \AA, 2365 \AA, and 2740 \AA\, \citep{1978csuf.book.....T}, and  spectra from the International Ultraviolet Explorer (IUE)\footnote{http://archive.stsci.edu/iue/} obtained through the large aperture. In the optical range, we used data from Adelman's spectrophotometric catalogue \citep{1989A&AS...81..221A}. 
Photometric data in the IR region were taken from the 2Micron All-Sky Survey \citep[2MASS,][]{2003yCat.2246....0C}, which contains an overview of the entire sky in J (1.25 $\mu$m), H (1.65 $\mu$m), and Ks (2.17 $\mu$m) filters. Observations were transformed into the absolute fluxes using the calibrations given in \citet{2003AJ....126.1090C}.

\section{Stellar fundamental parameters}
\label{model}
\subsection{\SME}
Fundamental parameters of sample stars were determined using \SME\, (Spectroscopy Made Easy) spectral package \citep{1996AAS..118..595V, 2017A&A...597A..16P}. It  
computes synthetic spectra in a given spectral regions taking into account all possible blends, performs fit to the observed spectra, and finds best-fit solution for the atmospheric parameters \teff, \lgg, apparent rotation velocity (\vs ), microturbulent velocity (\vmicro ), macroturbulent velocity (\vmacro ), and metallicity ([M/H]) within the grids of model atmospheres. \SME\, enables a careful selection of spectral features that exhibit sensitivity to different atmospheric parameters (line mask). These parameters are the free parameters of the fit. The present study is based on \textsc{LLmodels} grid \citep{LLmodels}. \SME\ masks are constructed in wide spectral regions 4250--6700~\AA\ (\Peg, \Vir, \Cap), 4400--6460~\AA\ (\Gem). 
Our masks include from three (H$\gamma$, H$\beta$, H$\alpha$) to one (H$\beta$) hydrogen lines, which are sensitive to \teff\ and, especially \lgg\ variations,  in the range of stellar parameters we are interested in.
The masks also contain $\approx$300 lines of Fe\ione\ and Fe\ii\ of different excitation energies with the equivalent widths (EW) from 3 to 110~m\AA\ in order to refine \teff\ and \lgg\ via excitation and ionisation equilibria as well as to determine \vmicro.

 The uncertainties of the free parameters are estimated with two methods: the first one is the classical method derived from the covariance matrix of 
the least square fitting procedure, while the second method is based on the analysis of cumulative probability distribution for each parameter
\citep[see][for more details]{2016MNRAS.456.1221R, 2017A&A...597A..16P, 2023A&A...671A.171W} and it includes the residuals of the fit that also 
reflect systematic errors such as detector defects, continuum normalisation, missing or erroneous atomic data, etc.
The distribution for each free parameter consists of the changes to this parameter that are needed to achieve a perfect fit (zero residual) in every spectral point. 
Only points sensitive to the parameter in question are included.
Central part of the distribution for each parameter is close to Gaussian with the uncertainty depending on the point statistics and it can be used to determine 
the median and the width of the distribution in the integral form (cumulative distribution). 
Every free parameter is treated independently and so this approach tends to overestimate the uncertainty if statistics is poorly sampled, which becomes the case when 
only a small number of spectral points in chosen regions is sensitive to the parameter. When statistics is good this method provides reasonable error estimates for free parameters that affect 
the majority of spectral points (e.g. \teff, [M/H]). The true values of uncertainties of the free parameters lies somewhere in between the two estimates.

\subsection{\SED}
To check the accuracy of the fundamental parameters determination, we used spectral energy distribution (\SED).
When calculating the \SED\ for all stars except \Cap, the value of \lgg\, was taken from the \SME\, solution and fixed in the fitting procedure. For \Cap\ we derived same temperatures with fixed and not fixed \lgg.
For each star, we constructed a set of observed fluxes using the data described in Section~\ref{sec1:obs}. For \Gem, we used the UV data from the IUE and TD1 missions, in the optical range we used Adelman's spectrophotometry, and 2MASS photometric measurements in the IR. For \Peg, \Vir, and \Cap, the photometric and spectrophotometric data were taken from the IUE, TD1, Adelman's, and 2MASS catalogues. 
Sample stars are nearby objects located within 100 pc, however, we accounted for the interstellar absorption in flux calculations. 
The correction for the interstellar reddening was applied according to extinction curve from \citet{1999PASP..111...63F} with $A_v = 3.1*E(B-V)$.
All values of $E(B-V)$ were taken from the dust map given by \citet{2014A&A...561A..91L}. The corresponding $A_v$ are 0.04, 0.01, 0.0 and 0.01 for \Peg, \Vir, \Gem, and \Cap, respectively.

\subsection{Results}
The fundamental parameters for all sample stars are presented in Table~\ref{Fund_param_AmStars}. For \SME\ parameters, uncertainties estimated by the two methods are given, while the \teff\ uncertainty from  \SED\ determination is obtained by the first classical method mentioned above.

\begin{table*}
    \caption{Fundamental parameters of the program stars derived in the present study.}
    \scriptsize
    \setlength\tabcolsep{2pt}
    \centering
    \begin{tabular}{l | l l l l l l l l l l}
	\hline     Star name & \teff, K      &          \lgg          & [Fe/H]                & \vmicro, \kms         &\vmacro, \kms          & \vs, \kms              &   $R/R_\odot$ & $L/L_\odot$   & Parallax, mas & Method\\
	\hline
	\Gem  & 9100$\pm$10         &  3.60 &&&&& 5.16$\pm$0.77 & 2.21$\pm$0.13 & 29.84*   & \SED\,\\
	      & 9190$\pm$20$\pm$130 & 3.56$\pm$0.01$\pm$0.08 &0.06$\pm$0.01$\pm$0.08 & 1.77$\pm$0.03$\pm$0.36& 5.43$\pm$0.65$\pm$3.93&10.59$\pm$0.23$\pm$1.71 &&&& \SME\\
	\hline
	\Peg  & 9550$\pm$10         &  3.80 &&&&& 3.37$\pm$0.10 & 1.93$\pm$0.03 & 11.65**  & \SED\,\\
	      & 9600$\pm$10$\pm$50 & 3.81$\pm$0.01$\pm$0.04 &0.25$\pm$0.01$\pm$0.09 & 1.98$\pm$0.04$\pm$0.32& 5.20$\pm$0.64$\pm$2.34&~5.40$\pm$0.49$\pm$1.11 &&&& \SME\\
	\hline
	\Vir  & 9660$\pm$20         &  3.60  &&&&& 4.03$\pm$0.30 & 2.10$\pm$0.07 & 11.18**  & \SED\,\\
	      & 9600$\pm$10$\pm$140 & 3.61$\pm$0.01$\pm$0.12 &0.15$\pm$0.01$\pm$0.14 & 1.42$\pm$0.01$\pm$0.62& 3.78$\pm$0.03$\pm$1.52&~0.10$\pm$2.26$\pm$7.80   &&&& \SME\\
	\hline
	\Cap  &10160$\pm$10         & 3.95$\pm$0.08 &&&&& 3.04$\pm$0.08 & 1.95$\pm$0.02 & 12.17**  & \SED\,\\
	      &10200$\pm$10$\pm$220 & 3.88$\pm$0.01$\pm$0.08 &0.08$\pm$0.01$\pm$0.14 & 0.96$\pm$0.02$\pm$0.55& 0.00 &22.85$\pm$0.04$\pm$2.56 &&&& \SME\\
	\hline
    \end{tabular}
    {\it Note.} *\citet{2007AA...474..653V}; **\citet{2020yCat.1350....0G}
    \label{Fund_param_AmStars}
\end{table*}

\begin{figure*}
    \centering
    \includegraphics[width=1.00\linewidth, clip]{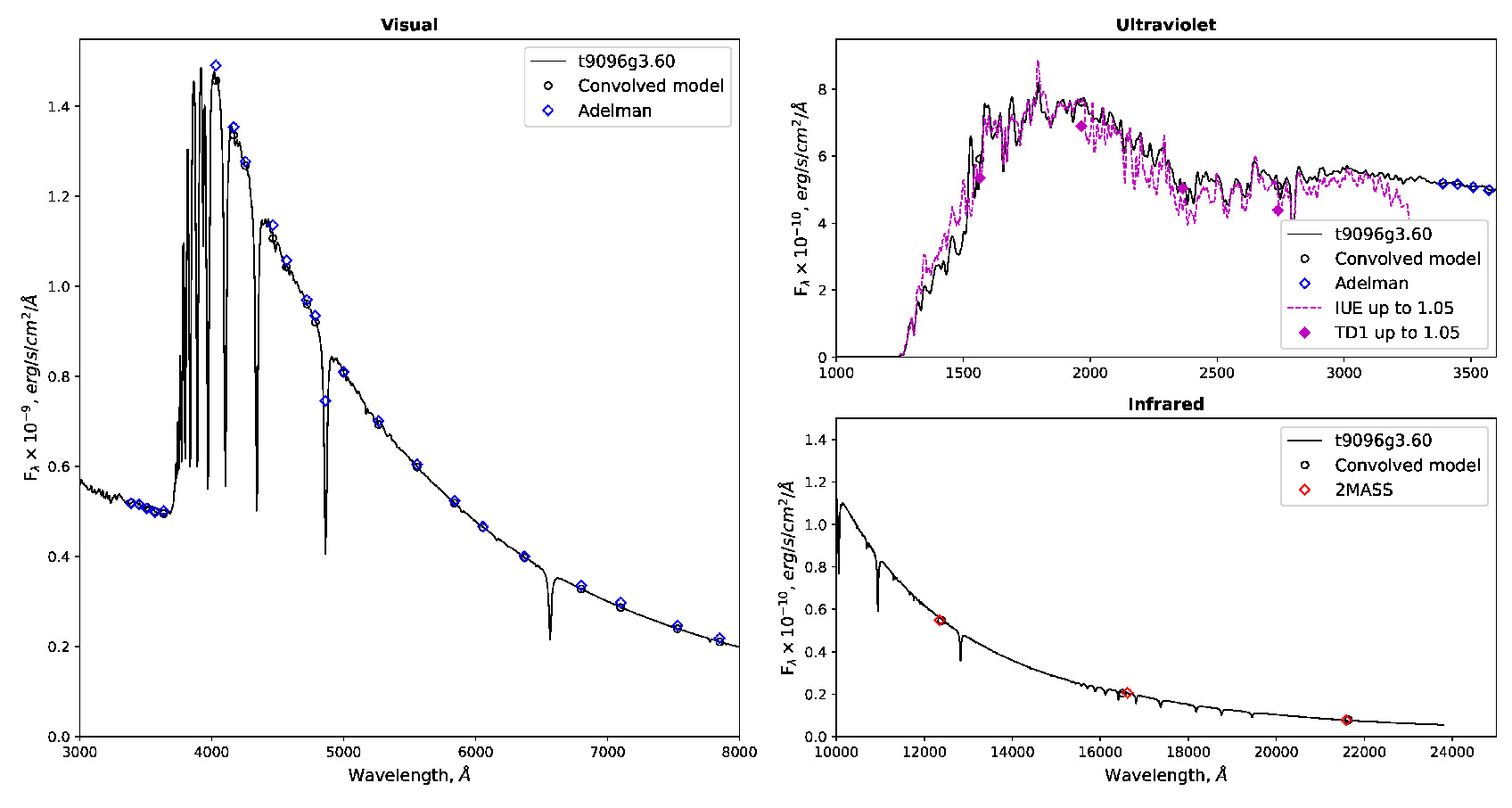}
    \caption{Spectral energy distribution of \Gem\, in the UV, visual, and IR spectral regions with Adelman's photometry data points. Black solid line shows theoretical \SED\, calculation with the model atmosphere parameters \teff\ =  9096~K, \lgg\ = 3.6. IUE and TD1 points were raised by a factor of 1.05.}
    \label{fig:gamgem-sed}
\end{figure*}

For \Gem, we obtained \teff\, = 9190 K and \lgg\, = 3.56 from \SME. From \SED\ we derived \teff\, = 9100~K (Fig.~\ref{fig:gamgem-sed}). However, as in the case of the magnetic star HD~108662 \citep[see][]{2020INASR...5..219R}, it turned out that there is a noticeable disagreement in the narrow spectral region between the UV and optical observations. A good agreement between the observed and theoretical energy distribution could only be obtained assuming a scaling factor of 1.05 for the IUE flux and TD1 photometry.

For \Peg, we obtained \teff\ = 9600~K, \lgg\ = 3.81 from \SME, and \teff\, = 9550~K from \SED\, (see Fig.~\ref{fig:omipeg-sed} in Appendix). In this case we keep assuming a scaling factor of 1.05 for the IUE flux and TD1 photometry with the same reasons as for \Gem.
  
For \Vir, atmospheric parameters from \SME\, are \teff\ = 9600~K, \lgg\ = 3.61 and \SED\, gives \teff\ = 9660~K (see Fig.~\ref{fig:thetavir-sed} in Appendix). 

For \Cap, we obtained \teff\ = 10200~K, \lgg\ = 3.88 from \SME\, analysis and \teff\ = 10150~K from \SED\, fitting (see Fig.~\ref{fig:nucap-sed} in Appendix).  

For each program stars, the atmospheric parameters derived from spectroscopic and spectrophotometric observations agree well within the error limits.

\begin{figure}
    \centering
    \includegraphics[width=1.0\linewidth, clip]{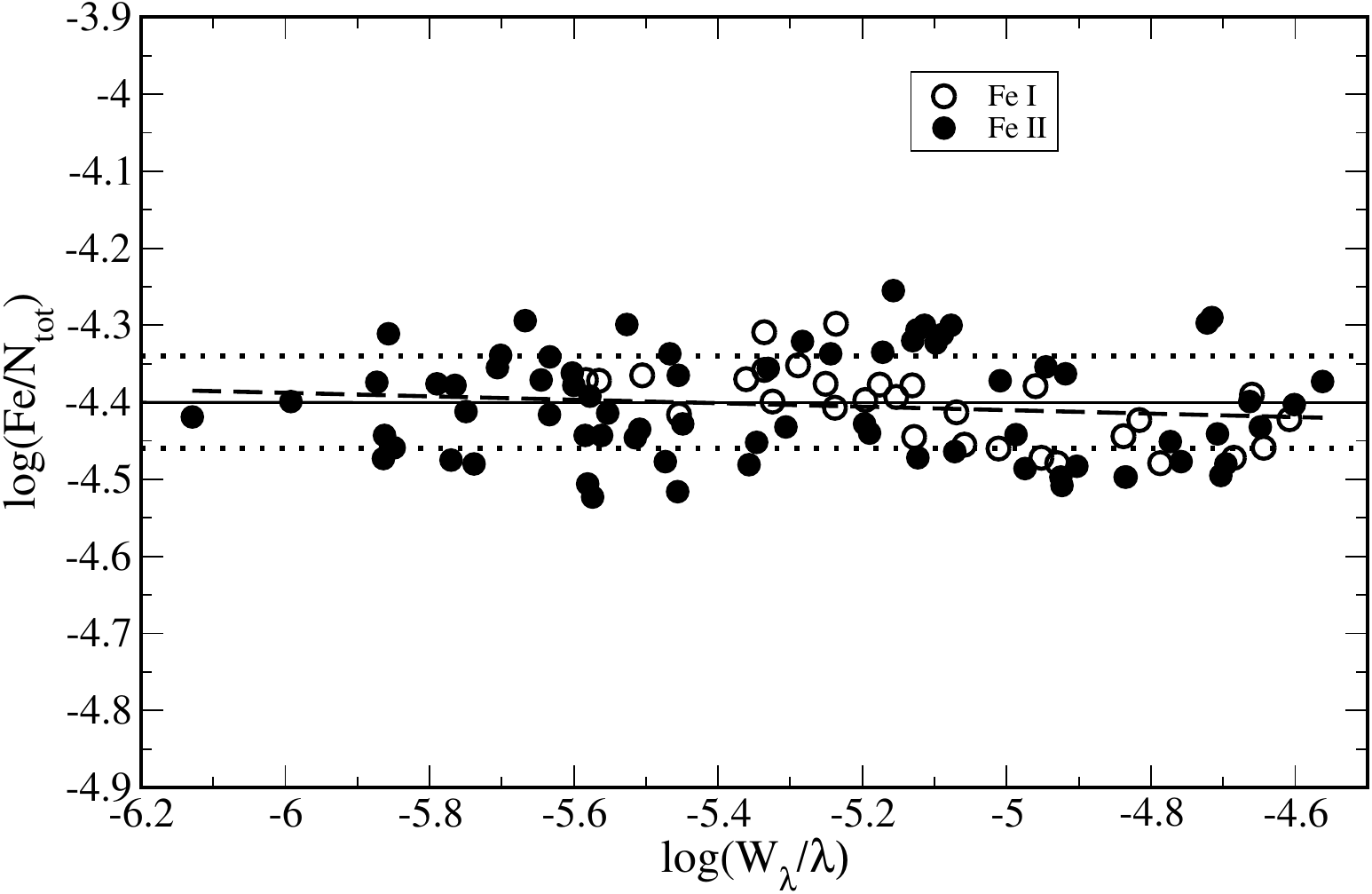}
    \caption{Abundances from Fe~\ione\ (open circles) and Fe\ii\ (filled circles) lines in \Gem\, versus the reduced equivalent widths. Solid and dashed lines represent an average Fe abundance and regression line of the entire set, correspondingly. The uncertainty limits are shown by dotted lines.}
    \label{fig:gamgem-vmic-ew}
\end{figure}

Along with the \teff\  and \lgg, the microturbulent velocity is another parameter that affects abundances. It is adjusted in a such way that, for a given species, spectral lines of different equivalent width yield similar abundances.  
We checked \vmicro\, obtained from \SME\, for \Gem\, (see Fig.~\ref{fig:gamgem-vmic-ew}) by manually fitting the individual spectral line profiles of Fe\ione\ and Fe\ii\, lines (see Section~\ref{abund}) and found that the \vmicro\ found in SME meets the above criteria.  
The same is true for \Peg, \Vir, and \Cap. 

\section{Abundance analysis}
\label{abund}
Abundance analysis was carried out with the model atmospheres obtained from \SME. For each chosen line, abundance is determined by fitting the synthetic line profile to the observed profile using the code \SV\, \citep{2019ASPC..518..247T} along with the code \bmag\,\citep{2018ascl.soft05015K}, which may implement the pre-computed departure coefficients (the ratio of NLTE to LTE atomic level populations). For elements up to Ba, the abundances were obtained from the lines of the neutral atoms and first ions, while for the rare-earth elements (REE), we used the lines of the first and second ions. 

Element abundances are given in a standard designation, \abun\ = $\log(N_{\rm El}/N_{\rm H})$ + 12, where N$_{\rm El}$ and N$_{\rm H}$ are number densities of a given chemical element and hydrogen, respectively. For each star, the full linelist with the line atomic parameters and individual line abundances is available online (see Section~\ref{dataavailability} for details). Part of the list, for guidance, is given in Table~\ref{tab:linelist}.

\begin{table*}
\caption{The linelist of LTE and NLTE abundances from individual lines in program stars together with the line atomic parameters and the references to them.}
\label{tab:linelist}
 \scriptsize
    \setlength\tabcolsep{2pt}
    \centering
    \begin{tabular}{lccc|ccc|cccccccc}
	\hline &&&&   \multicolumn{3}{c}{References}& \multicolumn{8}{c}{Abundances, \abun} \\ \cmidrule(lr){5-7}\cmidrule(lr){8-15}
	&&&&&&& \multicolumn{2}{c}{\Gem} & \multicolumn{2}{c}{\Peg} & \multicolumn{2}{c}{\Vir} & \multicolumn{2}{c}{\Cap} \\
	Ion & Wavelength, \AA & $E_i$, eV & log$gf$ & gf & HFS & IS & LTE & NLTE & LTE & NLTE & LTE & NLTE & LTE & NLTE \\
	\hline
	\multicolumn{15}{c}{. . .} \\
	He 1	& 4471.4730 &20.9641 &-0.2780 & WSG & - &- & 10.93 & 10.90 & 10.89 & 10.87 & 10.71 & 10.69 & 10.96 & 10.95 \\
	He 1 & 5875.6150 & 20.9641 & 0.4090 & WSG & - & - & 11.02 & 10.92 & 10.93 & 10.86 & 10.89 & 10.81 & 11.01 & 10.94 \\
	\multicolumn{15}{c}{. . .} \\
        Ba 2 & 4554.0319 & 0.0000 & 0.1700 & MW & BBW/VAHW & WABM & 2.64 & 2.92 & 3.44 & 3.62 & 3.19 & 3.49 & 2.76 & 3.09 \\
        Ba 2 & 4934.0750  & 0.0000 & -0.1500 & MW & BWE-BBW/VAHW & WABM & 2.65 & 2.94 & 3.40 & 3.61 & 3.18 & 3.49 & 2.79 & 3.11 \\
        Ba 2 & 5853.6742  & 0.6043 & -1.0000 & MW & VBDSb/VAHW & VBDS & 2.84 & 3.14 & 3.45 & 3.69 & 3.28 & 3.61 & - & - \\
        \multicolumn{15}{c}{. . .} \\
	\hline
    \end{tabular}
    
{\it Note.} This table is available in its entirety in a machine-readable form in the online journal. A portion is shown here for guidance regarding its form and content. References of HFS constants are given for lower and upper levels. WSG = \cite{1966atp..book.....W}; MW = \cite{1969AD......1....1M}; BBW = \cite{1981pmfc.conf...99B}; VAHW = \cite{1993JPhB...26.4289V}; WABM = \cite{1984ZPhyA.318..125W}; BWE = \cite{1982PhRvA..25.1476B}; VBDSb = \cite{1985ZPhyA.321..215V}; VBDS = \cite{1982JPhB...15.1805V}.
\end{table*}

For most chemical elements, we determined NLTE abundances, while Sc, V, Cr, Mn, Co, Ni, La, Nd abundances are determined in LTE.

 For C\ione\ --\ii, O\ione, Na\ione, Mg\ione\ --\ii, Si\ione\ --\ii, K\ione, Ca\ione\ --\ii, Ti\ione\ --\ii, Fe\ione\ --\ii, Zn\ione\ --\ii, Sr\ii, Y\ii, Zr\ii\ --\iii, and Ba\ii, the
NLTE calculations are performed with a modified version of the \textsc{DETAIL} code \citep{Giddings81,Butler84} with the opacity package  updated by \citet{2011JPhCS.328a2015P}. For He\ione, N\ione\ --\ii, Al\ione, and S\ione, NLTE calculations are performed with the \textsc{MULTI} code \citep{1986UppOR..33.....C} where the opacity calculation block was replaced with that from \textsc{ATLAS9} \citep{2003IAUS..210P.A20C} as described by \citet{1999AA...351..168K}. We are using two NLTE codes because model atoms for some species are created for a specific code. However, we have compared NLTE calculations with \textsc{DETAIL} and \textsc{MULTI} codes for common elements Na\ione\, and Mg\ione\ in \Gem. The derived abundance difference between two NLTE codes does not exceed 0.03~dex which is comparable to the  uncertainty of NLTE abundance determinations.

The NLTE abundances of He, C, O, Na, Mg, Si, Ca, Ti, Fe, Sr, Zr, and Ba are determined using the same NLTE methods
as in Paper~\ione\ and we refer the reader to that study. In addition to the list of chemical elements investigated in Paper~\ione, here we also determine NLTE abundances of N, Al, S, K, Zn, Y in the four sample stars and the comparison stars from Paper~\ione\ (HD~32115, 21~Peg, Sirius, and HD~72660). For these elements, their abundance determination methods are described below. 
The derived abundances for the sample stars are presented in Table~\ref{MeanAbunds}, while our results for  comparison stars are given in Table~\ref{MeanAbunds-21Peg-Sirius}

\underline{Nitrogen}
NLTE calculations were performed with the model atom originally described by \citet{2011MNRAS.410.1774L} and modified by \citet{2021AN....342..887A}. This model was supplemented by the detailed calculations of low-energy inelastic collisions with hydrogen atoms  \citep{2019AA...625A..78A}. Twenty so-called super levels of N\ione\ and transitions between them were added in order to reliably describe the relationship between the excited levels of neutral nitrogen and the continuum. Seventy-five highly excited N\ione\ levels were combined in 20 super levels according to their parity. In total, the model atom contains 59 N\ione\ levels, 49 N\ii\ levels, N\iii\ ground state, and more than 900 b-b transitions between them.

In \Gem\, and \Peg, we determined nitrogen NLTE abundances  \abun\ = 7.74$\pm$0.07 and  8.05$\pm$0.06, respectively. Our results agree with the NLTE determinations of \citet{2018PASJ...70...91T}, who found  \abun\ = 7.74 and 8.00, from analysis of the single N\ione\ 7468~\AA\ line in the same stars. 
For Sirius and \Cap, we found 0.23~dex and 0.18~dex lower abundances compared to those of \citet{2018PASJ...70...91T}. Both stars have the largest projected rotational velocities among the program and the comparison stars.
Abundance analysis of \citet{2018PASJ...70...91T} was based on EW measurements, therefore partially this difference may be attributed to a small blend of S~\ione\ 7868.59~\AA\ line, which may affect the total EW in a different way in stars with different rotational velocity. 

\underline{Aluminium}
The two resonance Al\ione\ lines and several Al\ii\ lines are suitable for abundance determination, but the resonance lines are strongly affected by NLTE. 
For NLTE calculations we used the Al\ione\ model atom described in details in \citet{2021AN....342..887A} and modified by \citet{2019AA...628A..46C}. This model atom consists of 76 levels of Al\ione\ and 13 levels of Al\ii\ and it takes into account about 300 b-b transitions between them. NLTE effects lead to a significant weakening of the resonance lines and  the NLTE corrections reach 0.40 dex in the stars under study. High-excitation Al\ii\ lines were considered in LTE approximation. For the four sample stars, the abundance difference between Al\ione\ and Al\ii\ varies from 0.03 to 0.17~dex in different stars. We consider this result to be satisfactory, due to possible uncertainties in NLTE calculations for Al\ione\ resonance lines caused by their position in the wing of the strong Ca\ii\ 3933~\AA\ resonance line.

\underline{Sulfur}
NLTE calculations for S\ione\ were performed with the model described in details in \citet{2009ARep...53..651K}. The model atom includes 64 levels of singlet, triplet and quintet systems of S\ione, the ground level of S\ii\ and about 400 radiative transitions between them. In our analysis, transition probabilities for S\ione\ and S\ii\ lines are taken from \citet{ZB}  and  \citet{IFF_2005}, respectively. NLTE corrections for the S\ione\ lines adopted for abundance determination in this study are small in the temperature domain of our sample stars. They can be either positive or negative and range from --0.04 to +0.10 dex.  However, the IR lines of the S\ione\ 9221-9237 \AA\ triplet are always strengthened due to NLTE effects, and their NLTE abundance corrections range from --0.5 to --0.8~dex. 

For S\ii\ lines, we performs LTE analysis. For the four stars where lines of S\ione\ and S\ii\ are both measured, we found  NLTE abundances from S\ione\ to be systematically higher compared to S\ii, with the abundance difference from 0.6 to 0.24~dex.   Partially, the difference of ~0.2~dex is probably caused by the uncertainty in theoretical transition probabilities, which is ~0.15~dex according to NIST \citep{NIST_ASD_2022}.

\underline{Potassium}  We apply a comprehensive model atom of K\ione\ developed by \citet{2020AstL...46..621N} using the most up-to-date atomic data on energy levels (46 levels with a principal quantum number $n \le 14$), transition probabilities, photoionisation cross-sections, and electron-impact excitation rates. Singly-ionised potassium is represented by the ground state only because the first excited level has an excitation energy of 20.2~eV. 

\underline{Zinc} 
NLTE calculations for Zn are performed with a Zn\ione\ --\ii\ model atom of \citet{2022MNRAS.515.1510S}. It includes 28 levels of Zn\ione, 7 levels of Zn\ii, and the ground state of Zn\iii. This model atom employs photoionisation cross-sections from the R-matrix calculations of \citet{2011AA...536A..51L} and electron impact excitation cross sections from the R-matrix calculations of \citet{2005PhRvA..71b2716Z}. In our sample stars, NLTE leads to higher Zn abundances compared to LTE. 

\underline{Yttrium} 
We construct a model atom for Y\ii\ using atomic level energies and
oscillator strengths from calculations of R. Kurucz available on his
web-page\footnote{http://kurucz.harvard.edu/atoms.html}. Our Y\ii\
model atom includes 94 levels of Y\ii\ and the ground state of Y\iii.
Energy levels with E$_{exc} >$ 10~eV are combined in 22 super levels
according to their parity.
For electron impact excitation and ionisation and photoionisation we
use approximate formulae to calculate the transition rates. Namely,
hydrogenic formula with an effective principal quantum number for
photoionisation, \citet{1962amp..conf..375S} formula for electron
impact ionisation, and \citet{Reg1962} and \citet{1948MNRAS.108..292W}
formulae for electron-impact excitation for radiatively allowed and
forbidden transitions, respectively. To estimate the accuracy of our
NLTE results, we perform test calculations with scaling electronic
collision rates and photoionisation cross sections by one order of
magnitude. For the investigated lines, an application of ten times
smaller photoionisation rates or ten times larger electron impact
ionisation rates does not affect the NLTE abundances, while ten times
increase in electron collision excitation rates results in
significantly smaller departures from LTE. For example, for Sirius,
the latter test calculations result in up to 0.1~dex lower NLTE
abundances compared to that derived with the standard rates. The
effect is different for different spectral lines: it is small for
Y\ii\ lines in the UV spectral range and reaches maximum for the
strongest lines in the visible spectrum.

 In atmospheres of A-type stars, Y\ii\ is the minority species.
For example, in Sirius, the number density of Y\ii\ is an order of
magnitude smaller compared to that of Y\iii\ in the line formation
layers. The mechanism of the departures from LTE is driven by an
overionisation and results in a higher abundance compared to LTE. This
effect increases with \teff. For the hottest sample stars, NLTE
results in up to 0.5 dex higher average abundance compared to LTE. An
exception is the coolest F-type star HD~32115, where Y\ii\ dominates
and has small negative NLTE abundance corrections.

\begin{table*}
        \caption{Average NLTE and LTE chemical element abundances in the program stars.} 
        \small
        \renewcommand{\arraystretch}{0.8}
        \centering
        \begin{tabular}{l c |ccr|ccr|ccr|ccr|c}
                \hline \hline Ion & & \multicolumn{3}{c}{\Gem} & \multicolumn{3}{c}{\Peg} & \multicolumn{3}{c}{\Vir} & \multicolumn{3}{c}{\Cap} & Sun\\ 
                \hline   & & \abun & [X/H] & n$_{\rm l}$ & \abun & [X/H] &  n$_{\rm l}$ &  \abun & [X/H] & n$_{\rm l}$ & \abun & [X/H] & n$_{\rm l}$ & \abun \\
                \hline 
                He\ione  & L & 10.94(05)& 0.016 &  4 &10.89(03) & -0.034 &  4 &10.80(08) & -0.124 &  3 & 10.98(03) & 0.056 & 4 & 10.924\\
                He\ione  & N & 10.90(01)& -0.024 &  4 &10.88(03) & -0.044 &  4 &10.76(05) & -0.164 &  3 & 10.94(01) & 0.016 & 4 & \\
                C\ione   & L & 8.50(09) & 0.03 &  9 & 8.08(12) & -0.39 & 11 & 8.28(13) & -0.19 &  8 &  8.30(11) &-0.17 & 4 & 8.47\\
                C\ione   & N & 8.48(08) & 0.01 &  9 & 8.13(10) & -0.34 & 11 & 8.35(09) & -0.12 &  8 & 8.41(09)  & -0.06 & 4 & \\
                N\ione   & L & 8.12(09) & 0.27 & 10 & 8.43(15) &  0.58 & 19 & 8.01(06) &  0.16 & 15 &  8.20(08) & 0.35 & 10 & 7.85\\
                N\ione   & N & 7.74(07) &-0.11 & 10 & 8.05(06) &  0.20 & 19 & 7.55(05) & -0.30 & 15 & 7.67(09)  & -0.18 & 10 & \\
                
                O\ione   & L & 9.13(53) &  0.40 & 15 & 8.86(44) & 0.13 & 18 & 8.99(53) &  0.26 & 12 & 8.76(09)  & 0.03 & 13 & 8.73\\
                O\ione   & N & 8.73(04) & 0.00 & 15 & 8.55(07) & -0.18 & 18 & 8.55(03) & -0.18 & 12 & 8.67(07)  &-0.06 & 13 & \\
                Na\ione  & L & 6.95(26) & 0.68 &  4 & 6.99(29) &  0.72 &  5 & 6.75(28) &  0.48 &  4 & 6.71(20)  & 0.44 &  3 & 6.27\\
                Na\ione  & N & 6.41(03) & 0.14 &  4 & 6.59(03) &  0.32 &  5 & 6.36(02) &  0.09 &  4 & 6.36(03)  & 0.09 &  3 & \\
                Mg\ione  & L & 7.79(19) & 0.27 &  6 & 7.62(14) &  0.10 &  7 & 7.67(20) &  0.15 & 11 & 7.67(13)  & 0.15 &  6 & 7.52\\
                Mg\ione  & N & 7.66(04) & 0.14 &  6 & 7.55(04) & 0.03  &  7 & 7.54(02) &  0.02 & 11 & 7.57(03)  & 0.05 &  6 & \\
                Mg\ii    & L & 7.65(14) & 0.13 &  6 & 7.61(17) &  0.09 &  9 & 7.53(10) &  0.01 & 10 & 7.57(14)  & 0.05 &  4 & \\ 
                Mg\ii    & N & 7.62(08) & 0.10 &  6 & 7.53(10) & 0.01  &  9 & 7.46(06) & -0.06 & 10 & 7.53(07)  & 0.01 &  4 & \\
    $\textnormal{[Mg/H]}_{mean}$ & N &  \multicolumn{3}{c}{0.12(06)}  &  \multicolumn{3}{c}{0.02(07)}  &  \multicolumn{3}{c}{-0.02(05)}  & \multicolumn{3}{c}{0.03(05)}  &\\
                Al\ione  & L & 6.17(05) &-0.25 &  2 & 6.51(05) &  0.09 &  2 & 6.32(02) & -0.10 &  2 & 6.15(11)  &-0.27 &  2 & 6.42\\
                Al\ione  & N & 6.54(02) & 0.12 &  2 & 6.90(04) &  0.48 &  2 & 6.65(01) &  0.23 &  2 & 6.44(09)  & 0.02 &  2 & \\
                Al\ii    & L & 6.46(09) & 0.04 &  3 & 6.80(04) &  0.38 &  4 & 6.48(10) &  0.06 &  3 &  6.41(07) &-0.01 &  3 & \\
    $\textnormal{[Al/H]}_{mean}$ & &  \multicolumn{3}{c}{0.12(07)}  &  \multicolumn{3}{c}{0.48(04)}  &  \multicolumn{3}{c}{0.23(01)}  &   \multicolumn{3}{c}{}  &\\
                Si\ione  & L & 7.29(23) &-0.22 &  2 & 7.35(18) & -0.16 &  2 & 7.44(09) & -0.07 &  2 &           &      &    & 7.51\\
                Si\ione  & N & 7.64(15) & 0.13 &  2 & 7.65(11) &  0.14 &  2 & 7.66(20) &  0.15 &  2 &   &  &  & \\ 
                Si\ii    & L & 7.65(15) & 0.15 & 14 & 7.78(17) &  0.27 & 14 & 7.57(19) &  0.06 & 12 & 7.66(19)  & 0.15 & 10 & \\ 
                Si\ii    & N & 7.54(10) & 0.03 & 14 & 7.66(10) &  0.15 & 14 & 7.47(11) & -0.04 & 12 & 7.49(08)  &-0.02 & 10 & \\ 
    $\textnormal{[Si/H]}_{mean}$ & N &  \multicolumn{3}{c}{0.08(12)}  &  \multicolumn{3}{c}{0.15(11)}  &  \multicolumn{3}{c}{0.06(16)}  &  \multicolumn{3}{c}{} &\\
                S\ione   & L & 7.39(08) & 0.24 &  7 & 7.66(05) &  0.51 & 11 & 7.55(08) &  0.40 &  7 &           &      &   & 7.15\\ 
                S\ione   & N & 7.36(08) & 0.21 &  7 & 7.63(05) &  0.48 & 11 & 7.52(07) &  0.37 &  7 &           &      &   & \\
                S\ii        & L &          &      &    & 7.48(08) &  0.33 & 10 & 7.29(06) &  0.14 &  6 &  7.18(06) & 0.03 & 6 & \\
    $\textnormal{[S/H]}_{mean}$ & &   \multicolumn{3}{c}{} &  \multicolumn{3}{c}{0.48(05)}  &  \multicolumn{3}{c}{0.37(07)}&   \multicolumn{3}{c}{} &\\
                K\ione   & L & 5.67(20) & 0.60 &  1 & 5.26(20) &  0.19 &  1 & 5.05(20) & -0.02 &  1 &   &  &  & 5.07\\
                K\ione   & N & 5.36(20) & 0.29 &  1 & 5.00(20) & -0.07 &  1 & 4.78(20) & -0.29 &  1 &   &  &  & \\
                Ca\ione  & L & 6.39(10) & 0.12 & 13 & 6.42(13) &  0.15 & 16 & 6.51(09) &  0.24 & 11 &  6.12(20) &-0.15 & 1 & 6.27\\
                Ca\ione  & N & 6.53(09) & 0.26 & 13 & 6.57(13) &  0.30 & 16 & 6.73(06) &  0.46 & 11 & 6.42(20)  & 0.15 & 1 & \\
                Ca\ii    & L & 6.35(04) & 0.08 &  7 & 6.41(06) &  0.014 & 11 & 6.49(11) &  0.23 & 13 &  6.18(03) &-0.09 & 6 & \\ 
                Ca\ii    & N & 6.44(04) & 0.17 &  7 & 6.48(04) &  0.21 & 11 & 6.62(08) &  0.35 & 13 & 6.50(03)  & 0.23 & 6 & \\ 
    $\textnormal{[Ca/H]}_{mean}$ & N &  \multicolumn{3}{c}{0.21(06)}  &  \multicolumn{3}{c}{0.25(08)}  &  \multicolumn{3}{c}{0.41(07)}  & \multicolumn{3}{c}{0.19(02)}  &\\
    
                Sc\ii    & L & 3.01(08) &-0.03 & 12 & 2.94(13) & -0.10 & 12 & 3.17(05) &  0.13 & 12 & 2.84(08)  &-0.30 & 7 & 3.04\\
                Ti\ii    & L & 5.11(08) & 0.21 & 45 & 5.16(12) &  0.26 & 49 & 5.26(10) &  0.36 &  46& 4.91(05)  & 0.01 & 30 & 4.90\\
                Ti\ii    & N & 5.07(06) & 0.17 & 45 & 5.13(12) &  0.23 & 49 & 5.23(09) &  0.33 &  46& 4.89(06)  &-0.01 & 30 & \\ 
                V\ii     & L & 4.14(08) & 0.19 &  6 & 4.46(10) &  0.51 &  7 & 4.45(03) &  0.50 &  7 & 4.12(06)  & 0.17 & 5 & 3.95\\
                Cr\ione  & L & 5.85(12) & 0.22 &  8 & 5.95(10) &  0.32 & 10 & 5.81(05) &  0.18 &  6  & 5.88(06)  & 0.25 & 6 & 5.63\\
                Cr\ii    & L & 5.80(15) & 0.17 & 27 & 5.96(14) &  0.33 & 29 & 5.84(11) &  0.21 & 33 & 5.85(07) & 0.22 & 32 & \\ 
    $\textnormal{[Cr/H]}_{mean}$ & L &  \multicolumn{3}{c}{0.19(14)}  &  \multicolumn{3}{c}{0.33(12)}  &  \multicolumn{3}{c}{0.19(08)}  & \multicolumn{3}{c}{0.23(06)}  &\\
                Mn\ione  & L & 5.52(10) & 0.05 &  8 & 5.70(08) &  0.23 &  8 & 5.51(06) & 0.04 &  8 & 5.35(04) &-0.12 & 3 & 5.47\\
                Mn\ii    & L & 5.52(06) & 0.05 &  5 & 5.81(09) &  0.34 &  6 & 5.60(11) &  0.13 &  5 & 5.46(04) &-0.01 & 2 &  \\ 
    $\textnormal{[Mn/H]}_{mean}$ & L &  \multicolumn{3}{c}{0.05(08)}  &  \multicolumn{3}{c}{0.29(08)} &  \multicolumn{3}{c}{0.09(08)}  & \multicolumn{3}{c}{-0.07(04)}  &\\
                Fe\ione  & L & 7.56(05) &  0.11 & 30 & 7.70(06) & 0.25 & 31 & 7.58(05) &  0.13 & 29 &  7.50(06) & 0.05 & 24 & 7.45\\
                Fe\ione  & N & 7.64(05) &  0.19 & 30 & 7.78(05) & 0.33 & 31 & 7.68(06) &  0.23 & 29 &  7.58(06) & 0.13 & 24 & \\ 
                Fe\ii    & L & 7.61(06) &  0.16 & 74 & 7.81(05) & 0.36 & 73 & 7.84(07) &  0.39 & 76 & 7.55(05) & 0.10 & 67 &  \\ 
                Fe\ii    & N & 7.60(06) &  0.15 & 74 & 7.80(05) & 0.35 & 73 & 7.83(07) &  0.38 & 76 & 7.56(05) & 0.11 & 67 &  \\ 
    $\textnormal{[Fe/H]}_{mean}$ & N &  \multicolumn{3}{c}{0.17(06)}  &  \multicolumn{3}{c}{0.34(05)}  &  \multicolumn{3}{c}{0.31(06)}  & \multicolumn{3}{c}{0.12(06)}  &\\    
                Co\ione  & L & 4.99(20) &  0.13 &  1 & 5.48(20) & 0.62 &  1 & 5.36(20) &  0.50 &  1 &   &  &  & 4.86\\
                Co\ii    & L & 5.12(08) &  0.26 &  4 & 5.74(02) & 0.88 &  4 & 5.46(06) &  0.60 &  3 &  &  &  & \\
    $\textnormal{[Co/H]}_{mean}$ & L &  \multicolumn{3}{c}{0.20(04)}  &  \multicolumn{3}{c}{0.75(01)}  &  \multicolumn{3}{c}{0.55(03)}  & \multicolumn{3}{c}{}  &\\
                Ni\ione  & L & 6.39(08) &  0.19 & 17 & 6.90(07) & 0.70 & 22 & 6.64(07) &  0.44 & 17 & 6.37(20)  & 0.17 & 1 & 6.20\\
                Ni\ii    & L & 6.45(05) &  0.25 &   7 & 7.08(09) & 0.88 &   9 & 6.76(09) &  0.56 & 10 & 6.40(03)  & 0.20 & 3 & \\
    $\textnormal{[Ni/H]}_{mean}$ & L &  \multicolumn{3}{c}{0.22(06)}  &  \multicolumn{3}{c}{0.79(08)}  &  \multicolumn{3}{c}{0.50(08)}  & \multicolumn{3}{c}{0.19(02)}  &\\
    
               Zn\ione   & L & 4.67(01) &  0.06 &  3 & 5.52(01) & 0.91 &  3 & 5.22(03) &  0.61 &  3 & 4.93(20)  & 0.32 & 1 & 4.61\\
               Zn\ione   & N & 4.84(02) &  0.23 &  3 & 5.68(01) & 1.07 &  3 & 5.37(03) &  0.76 &  3 & 5.07(20) & 0.46 & 1 & \\
                Sr\ii    & L & 2.74(03) & -0.14 &  2 & 3.80(05) & 0.92 &  3 & 3.62(05) &  0.74 &  4 & 3.05(07)  & 0.17 & 2 & 2.88\\
                Sr\ii    & N & 3.24(02) &  0.36 &  2 & 4.19(08) & 1.31 &  3 & 4.09(07) &  1.21 &  4 & 3.60(09)  & 0.72 & 2 & \\  
                Y\ii     & L  & 2.28(16) &  0.13 &  4 & 3.07(10) & 0.92 &  7 & 2.92(11) &  0.77 & 15 & 2.44(07)  & 0.29 & 3 & 2.15\\ 
                Y\ii     & N & 2.77(07) &  0.62 &  4 & 3.53(05) & 1.38 &  7 & 3.44(09) &  1.29 & 15 & 3.09(13)  & 0.94 & 3 & \\
                Zr\ii    & L & 2.81(12) &  0.26 &  6 & 3.58(10) & 1.03 &  7 & 3.38(09) &  0.83 &  6 & 2.70(01)  & 0.15 & 2 & 2.55\\
                Zr\ii    & N & 2.99(11) &  0.44 &  6 & 3.78(13) & 1.23 &  7 & 3.67(06) &  1.12 &  6 & 3.11(01)  & 0.56 & 2 & \\
                Ba\ii    & L & 2.71(08) &  0.54 &  4 & 3.45(03) & 1.28 &  5 & 3.22(07) &  1.05 &  6 &  2.82(07) & 0.65 & 3 & 2.17\\
                Ba\ii    & N & 3.00(09) &  0.83 &  4 & 3.66(04) & 1.49 &  5 & 3.55(06) & 1.38 &  6 & 3.15(07)  & 0.98 & 3 & \\
                La\ii    & L & 1.72(15) &  0.55 &  2 & 2.38(04) & 1.21 &  3 & 2.17(03) &  1.00 &  3 &   &  &  & 1.17\\
                Nd\iii   & L & 2.06(07) &  0.61 &  3 & 2.78(10) & 1.33 &  6 & 2.29(06) &  0.84 &  6 & 2.42(04)  & 0.97 & 2 & 1.42\\
                \hline \\
        \end{tabular}\\
        
	{\it Note.} L and N symbols indicate LTE and NLTE mean abundances. n$_{\rm l}$  is a number of spectral lines used for abundance determination.  
	The standard deviation is given in parentheses, and was assumed to be 0.2 dex when the measurement was obtained from a single line. The last column 
	contains present-day solar system abundances taken from \citet{2021SSRv..217...44L}. 
        \label{MeanAbunds}
\end{table*}

\section{Abundance comparison}\label{Comp1}
\begin{figure}
    \centering
    \includegraphics[width=1.0\linewidth, clip]{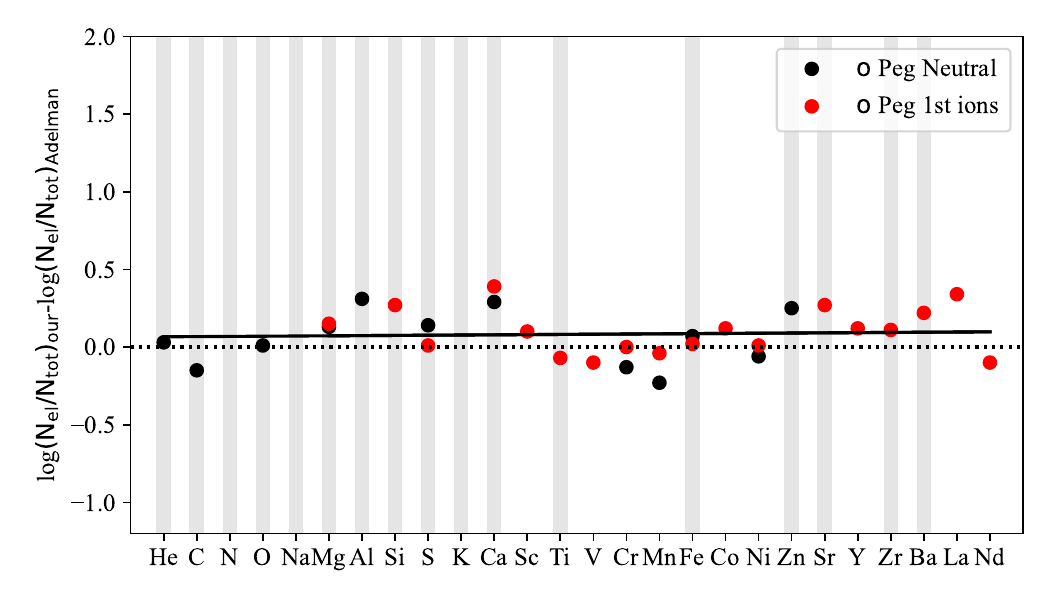}
    \caption{The difference of abundances derived in this study and those from \citet{2015PASP..127...58A} for \Peg\. Linear regression is shown by the solid line. Shaded stripes indicate chemical elements, for which we determine abundances in NLTE.}
    \label{oPeg-Ad}
\end{figure}

First, we compare the results of our abundance analysis with the most detailed published abundances: \citet{2015PASP..127..340A} for \Gem, \citet{2015PASP..127...58A} for \Peg, \citet{1987Afz....26...55D} for \Vir, and \citet{1991MNRAS.252..116A} for \Cap. An averaged over the all measured elements abundance differences between our determinations and those from the cited papers are: 0.07$\pm$0.19~dex (\Gem), 0.08$\pm$0.16~dex (\Peg), 0.09$\pm$0.22~dex (\Vir), and 0.14$\pm$0.20~dex (\Cap). Fig.~\ref{oPeg-Ad} shows the abundance differences for \Peg. 

The differences between our results and the literature data are partially caused by taking into account NLTE effects. Using different stellar atmosphere parameters and different transition probabilities also contributes to the discrepancy. While the average abundance difference of  $\sim$0.1~dex does not exceed the abundance uncertainties, it is worth noting that, for individual elements, the abundance difference reaches $\sim$0.3~dex, as in the case of calcium in \Peg. 

\begin{table*}
        \caption{Average abundances in the comparison stars Sirius, HD~72660, 21 Peg and HD~32115. } 
        \small
       \begin{tabular}{l|rr|rr|rr|rr|c}
                \hline \hline Ion & \multicolumn{2}{c}{Sirius} &\multicolumn{2}{c}{HD~72660} & \multicolumn{2}{c}{21 Peg} &\multicolumn{2}{c}{HD~32115} & Refs.\\ 
                \hline      & $\log\varepsilon$ & [X/H] & $\log\varepsilon$ &[X/H]& $\log\varepsilon$ &[X/H]& $\log\varepsilon$ &[X/H]& \\
                \hline 
          \bf{He*}  &\bf{10.82(03)} & \bf{-0.10} & \bf{10.72(04)} &\bf{ -0.20}& \bf{10.90(02)} &\bf{ -0.02}&                &            & (1) \\
           \bf{C* }  &\bf{ 7.71(14)} & \bf{-0.76} & \bf{ 8.02(08)} &\bf{ -0.45}& \bf{ 8.38(09)} &\bf{ -0.09}& \bf{ 8.45(08)} &\bf{ -0.02} & (2) \\        
                N*    &    ~7.75(07)  &     -0.10  &      7.32(08)  &     -0.53 &      7.53(07)  &     -0.32 &      7.96(01)  &      0.11  & (4) \\                
\bf{O* }  &\bf{ 8.44(05)} & \bf{-0.29} & \bf{ 8.43(07)} &\bf{ -0.30}& \bf{ 8.63(02)} &\bf{ -0.10}& \bf{ 8.86(11)} &\bf{  0.13} & (2) \\ 
\bf{Na*}  &\bf{ 6.54(02)} & \bf{ 0.27} & \bf{ 6.66(02)} &\bf{  0.39}& \bf{ 6.24(01)} &\bf{ -0.03}& \bf{ 6.17(07)} &\bf{ -0.10} & (2) \\
\bf{Mg*}  &\bf{ 7.51(06)} & \bf{-0.01} & \bf{ 7.77(06)} &\bf{  0.25}& \bf{ 7.50(04)} &\bf{ -0.02}& \bf{ 7.58(06)} &\bf{  0.06} & (2) \\
                Al*   &     6.34(01)  &     -0.08  &      6.98(08)  &      0.56 &      6.29(04)  &     -0.13 &      6.25(11)  &     -0.17  & (4) \\ 
\bf{Si*}  &\bf{ 7.72(11)} & \bf{ 0.21} & \bf{ 7.83(09)} &\bf{  0.32}& \bf{ 7.50(13)} &\bf{ -0.01}& \bf{ 7.59(15)} &\bf{  0.08} & (2) \\ 
                S\ione*&    7.59(12)  &      0.44  &                &           &      7.46(05)  &      0.31 &      7.22(06)  &      0.07  & (4) \\                        
                S\ii  &     7.53(20)  &      0.38  &      7.37(12)  &      0.22 &      7.24(14)  &      0.09 &                &            & (4) \\   
                K*    &               &            &      5.34(20)  &      0.27 &                &           &      5.00(20)  &     -0.07  & (4) \\                       
\bf{Ca*}  &\bf{ 6.09(05)} & \bf{-0.18} & \bf{ 6.62(09)} &\bf{  0.35}& \bf{ 6.30(15)} &\bf{ 0.03}& \bf{ 6.39(09)} & \bf{  0.12}& (2) \\
\bf{Sc}   &\bf{ 1.99(11)} & \bf{-1.05} & \bf{ 2.63(05)} &\bf{ -0.41}& \bf{ 2.60(06)} &\bf{ -0.44}& \bf{ 3.22(10)} & \bf{  0.18}& (2) \\
\bf{Ti*}  &\bf{ 5.15(04)} & \bf{ 0.25} & \bf{ 5.45(08)} &\bf{  0.55}& \bf{ 4.80(04)} &\bf{ -0.10}& \bf{ 4.76(05)} & \bf{ -0.14}& (2) \\ 
                V     &     4.50(04)  &      0.55  &      4.64(06)  &      0.69 &      3.93(07)  &     -0.02 &      3.93(11)  &      -0.02 & (4) \\ 
                Cr    &     6.24(08)  &      0.61  &      6.17(11)  &      0.54 &      5.61(06)  &     -0.02 &      5.77(12)  &       0.14 & (4) \\ 
                Mn    &     5.76(01)  &      0.29  &      5.86(08)  &      0.39 &      5.42(13)  &     -0.05 &      5.30(15)  &      -0.17 & (4) \\
\bf{Fe*}  &\bf{ 7.98(06)} & \bf{ 0.53} & \bf{ 8.10(16)} &\bf{  0.65}& \bf{ 7.51(07)} &\bf{  0.06}& \bf{ 7.55(11)} &\bf{  0.10} & (2) \\
                Co    &     5.73(20)  &      0.87  &      5.78(07)  &      0.92 &      5.07(07)  &      0.21 &      4.70(04)  &      -0.16 & (4) \\
                Ni    &     6.91(07)  &      0.72  &      7.01(07)  &      0.81 &      6.32(08)  &      0.12 &      6.07(09)  &      -0.13 & (4) \\ 
\bf{Zn*}  &\bf{ 5.58(02)} & \bf{ 0.97} & \bf{ 5.61(04)} &\bf{  1.00}& \bf{ 4.98(20)} &\bf{  0.37}& \bf{ 4.34(01)} &\bf{ -0.27} & (3) \\
\bf{Sr*}  &\bf{ 3.83(04)} & \bf{ 0.95} & \bf{ 4.35(04)} &\bf{  1.47}& \bf{ 3.49(02)} &\bf{  0.61}& \bf{ 3.28(04)} &\bf{  0.40} & (2) \\ 
                Y*    &     3.37(08)  &      1.22  &      3.52(03)  &      1.37 &      2.92(14)  &      0.77 &      2.28(11)  &      0.13  & (4) \\
\bf{Zr*}  &\bf{ 3.40(13)} & \bf{ 0.85} & \bf{ 3.85(11)} &\bf{  1.30}& \bf{ 2.92(10)} &\bf{  0.37}& \bf{ 2.82(03)} &\bf{  0.27} & (2) \\
\bf{Ba*}  &\bf{ 3.74(06)} & \bf{ 1.57} & \bf{ 3.67(03)} &\bf{  1.50}& \bf{ 3.16(05)} &\bf{  0.99}& \bf{ 2.47(07)} &\bf{  0.30} & (2) \\
\bf{Nd}  &\bf{ 2.94(03)} & \bf{ 1.52} & \bf{ 2.74(04)} &\bf{  1.32}& \bf{ 1.96(01)} &\bf{  0.54}& \bf{ 1.31(01)} &\bf{ -0.12} & (2) \\ 
                \hline 
\teff                 & \multicolumn{2}{c}{9850}& \multicolumn{2}{c}{9700}& \multicolumn{2}{c}{10400}& \multicolumn{2}{c}{7250}&(2)\\
\lgg                  & \multicolumn{2}{c}{4.30}& \multicolumn{2}{c}{4.10}& \multicolumn{2}{c}{3.55} & \multicolumn{2}{c}{4.20}&(2)\\  
                \hline 
\end{tabular} \\
        {\it Note.} Elements with NLTE abundance determinations are marked by *. The standard deviation is given in parentheses. Abundances taken from our earlier studies are shown by boldface. References are 
        (1) -- \citet{2018AstL...44..621K}; 
        (2) -- \citet{2020MNRAS.499.3706M}; 
        (3) -- \citet{2022MNRAS.515.1510S};
        (4) -- this study.
        \label{MeanAbunds-21Peg-Sirius}
\end{table*}

\begin{figure*}
    \begin{subfigure}[t]{0.49\textwidth}
         \centering
         \includegraphics[width=\textwidth]{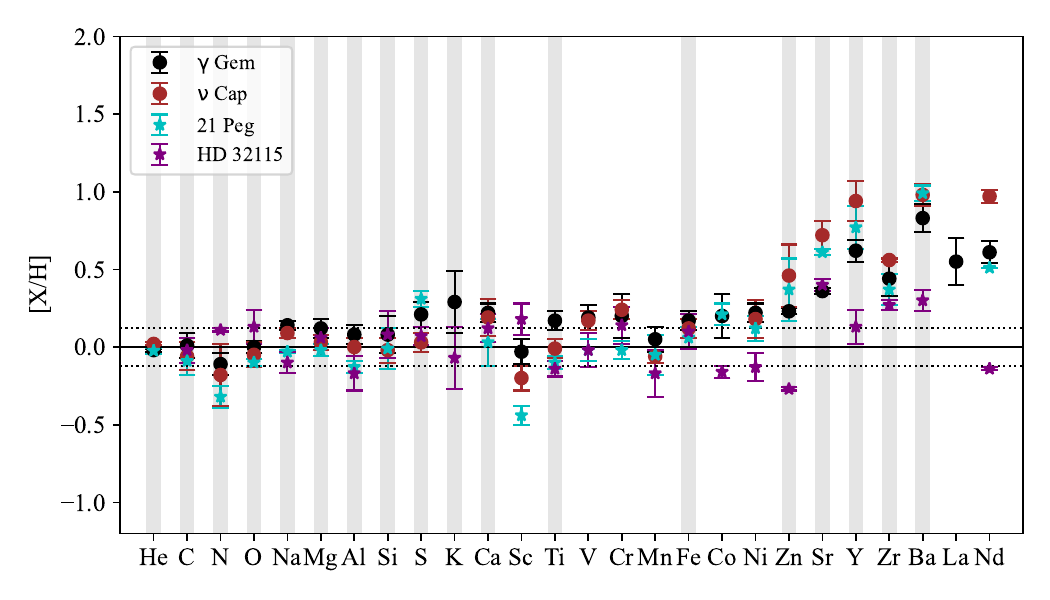}
        \caption{}
         \label{fig:gg-vs-21peg}
     \end{subfigure}
     \begin{subfigure}[t]{0.49\textwidth}
         \centering
         \includegraphics[width=\textwidth]{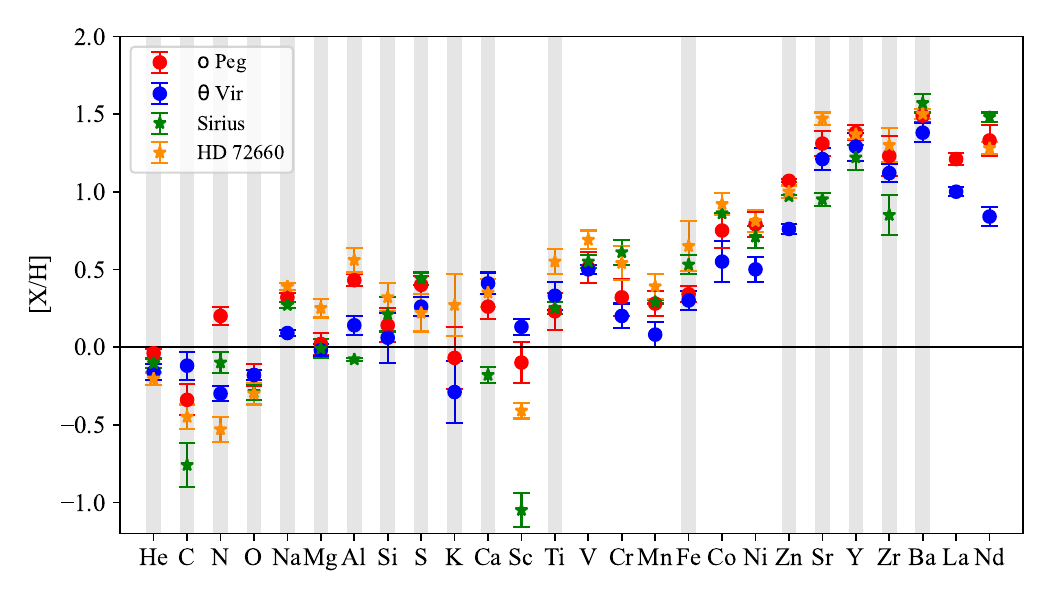}
        \caption{}
         \label{fig:op-tv-vs-sirius}
     \end{subfigure}
    \begin{subfigure}[t]{0.49\textwidth}
         \centering
         \includegraphics[width=\textwidth]{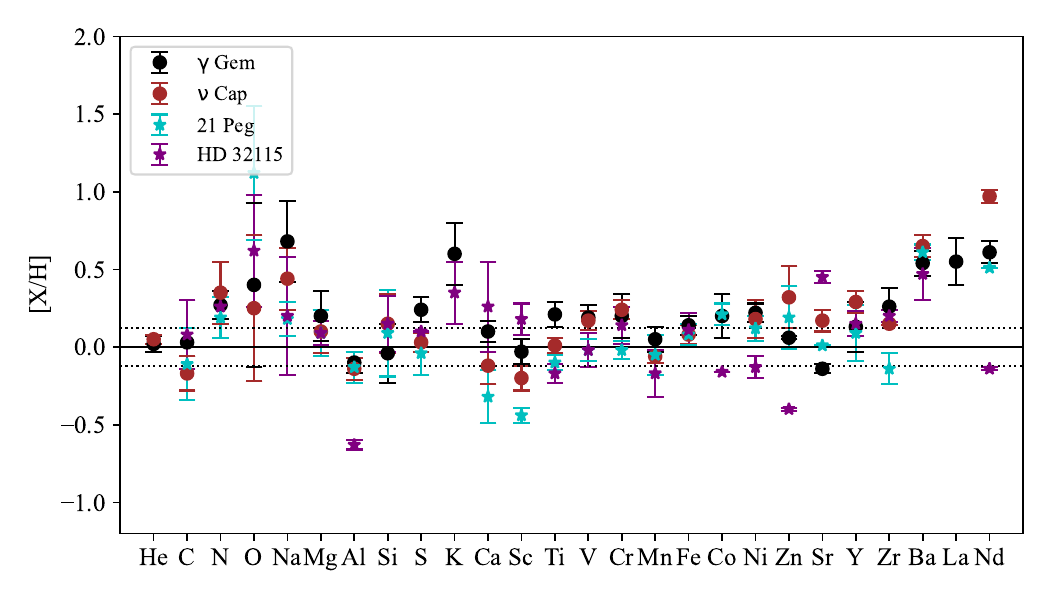}
        \caption{}
         \label{fig:gg-vs-21peg_LTE}
     \end{subfigure}
     \begin{subfigure}[t]{0.49\textwidth}
         \centering
         \includegraphics[width=\textwidth]{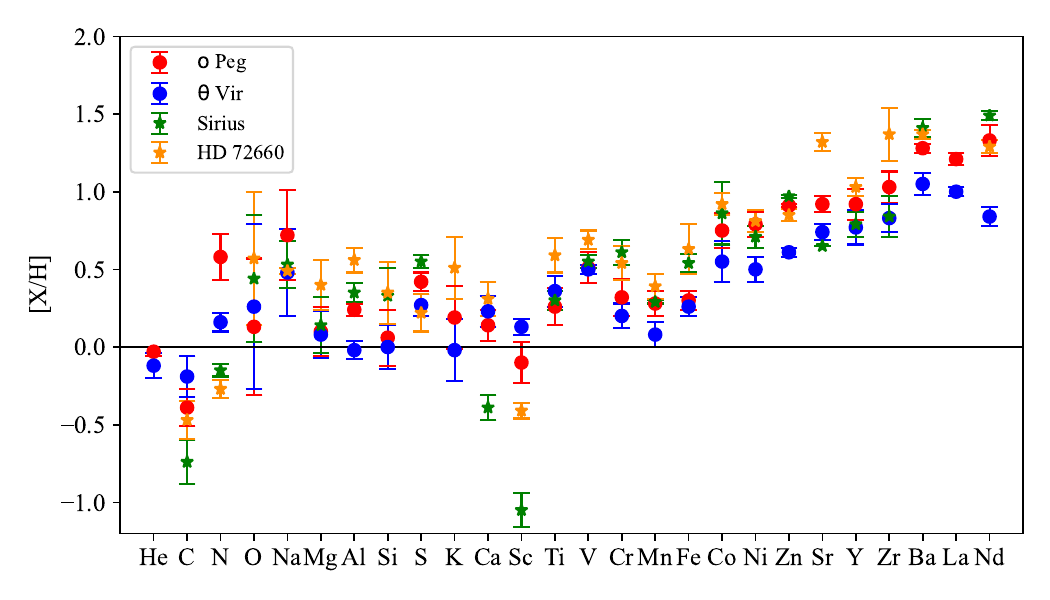}
        \caption{}
         \label{fig:op-tv-vs-sirius_LTE}
     \end{subfigure}
     \caption{Sample stars abundances in comparison with the abundances in normal stars HD~32115 and 21 Peg (a) and Am stars Sirius and HD~72660 (b). The dotted lines	 indicate 3$\sigma$ ($\pm$0.12~dex)  abundance error, which is typical for solar abundance analysis. Shaded stripes indicate chemical elements, for which we determine abundances in NLTE. Bottom panels (c) and (d) show LTE results for the same stars.}
\label{fig:abund}
\end{figure*}

The derived abundances for the sample stars and the comparison stars are shown in Fig.~\ref{fig:abund}. We split stars in two groups: normal A stars (\Gem, \Peg, 21~Peg, and HD~32115, Fig.~\ref{fig:gg-vs-21peg}) and Am stars (\Peg, \Vir, Sirius, and HD~72660, Fig.~\ref{fig:op-tv-vs-sirius}). NLTE results are displayed in the top panels, while pure LTE abundances are displayed in the bottom panels (Fig.~\ref{fig:gg-vs-21peg_LTE}, Fig.~\ref{fig:op-tv-vs-sirius_LTE}). For each group of stars, NLTE leads to smaller abundance differences between different stars. This effect is the most pronounced for light elements.

\subsection{Normal stars}
A star can be considered normal if its abundances for elements from He to Ni match solar values, with differences within $\pm$0.15~dex. 
Our detailed NLTE analysis supports the classification of both \Gem\, and \Cap\, as normal (superficially normal) stars. In \Gem\ and  \Cap,  abundances of all  elements up to Fe, except for N and Sc,  agree within $\pm$0.12~dex with the abundances in HD~32115 and 21~Peg, which are classified as normal stars.
Nitrogen deficiency in \Cap\, and in 21~Peg may be due to an overestimation of the NLTE effects. Scandium abundance was derived using Sc\ii\ lines, and abundance diversity may be caused by neglecting the NLTE effects. The  NLTE abundance corrections for Sc\ii\ lines were discussed in \citet{2020MNRAS.499.3706M}. They are expected to be positive for our sample stars.  

The elements heavier than Zn show overabundances of up to 1~dex. The possible correlation between the neutron capture elements (Sr-Zr-Ba-Nd) excess and effective temperature was noticed in Paper \ione. Here we refine this guess with the larger number of normal stars. The effective temperatures of our four normal sharp-lined stars range from 7200 to 10400~K. To extend this range towards higher \teff, we included Sr abunndance in the normal B-type star $\pi$~Cet (\teff\ = 12800~K) from Paper \ione. No lines of other heavy elements were detected in this star. 

\begin{figure*}
         \centering
         \includegraphics[width=\textwidth]{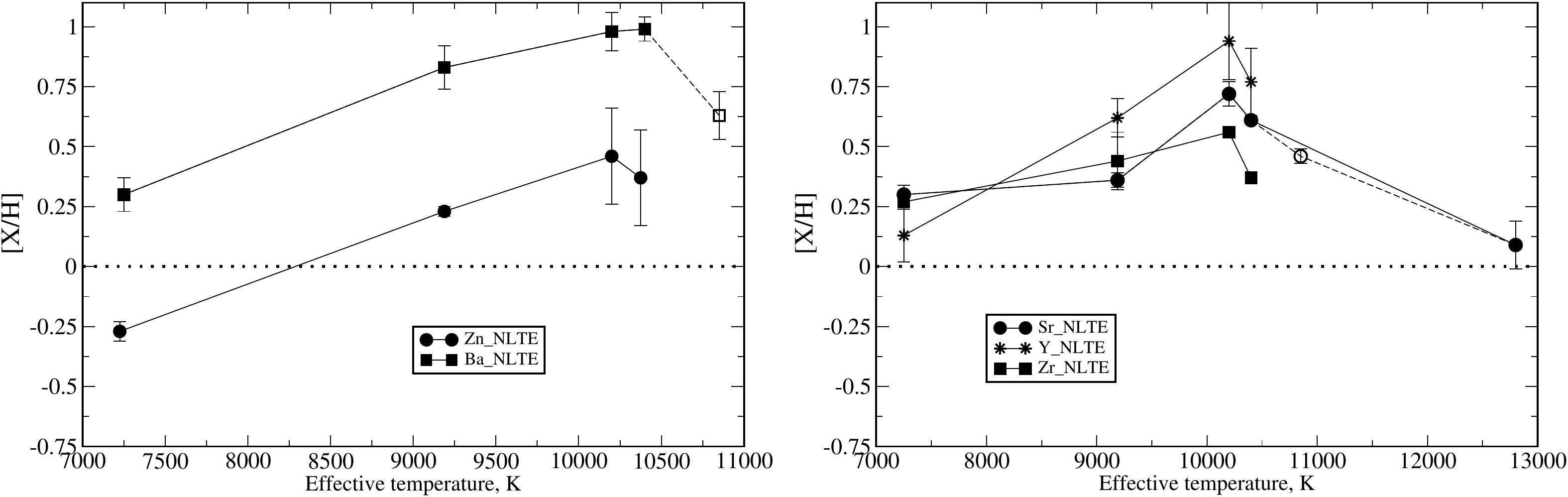}
     \caption{Neutron capture element abundance ratios in normal A stars as a function of \teff. Abundance ratios in 134~Tau are shown by the open symbols.}
         \label{Normal-Teff}
\end{figure*}

Fig.~\ref{Normal-Teff} shows an excess of heavy elements Zn-Ba-Sr-Y-Zr as a function of the effective temperature. A gradual increase in [X/H] of up to +1~dex is observed as \teff\ increases from 7200 to 10000~K. With a further increase in \teff, a hint of a decreasing trend can be noticed, and the Sr excess mostly reaches zero (solar abundance) at \teff\ = 12800~K. To support our present result, abundance studies of more stars with slow rotation and solar metallicity in temperature ranges 8000--9000~K and 9500-12000~K are required.

\begin{figure}
         \centering
         \includegraphics[width=0.49\textwidth]{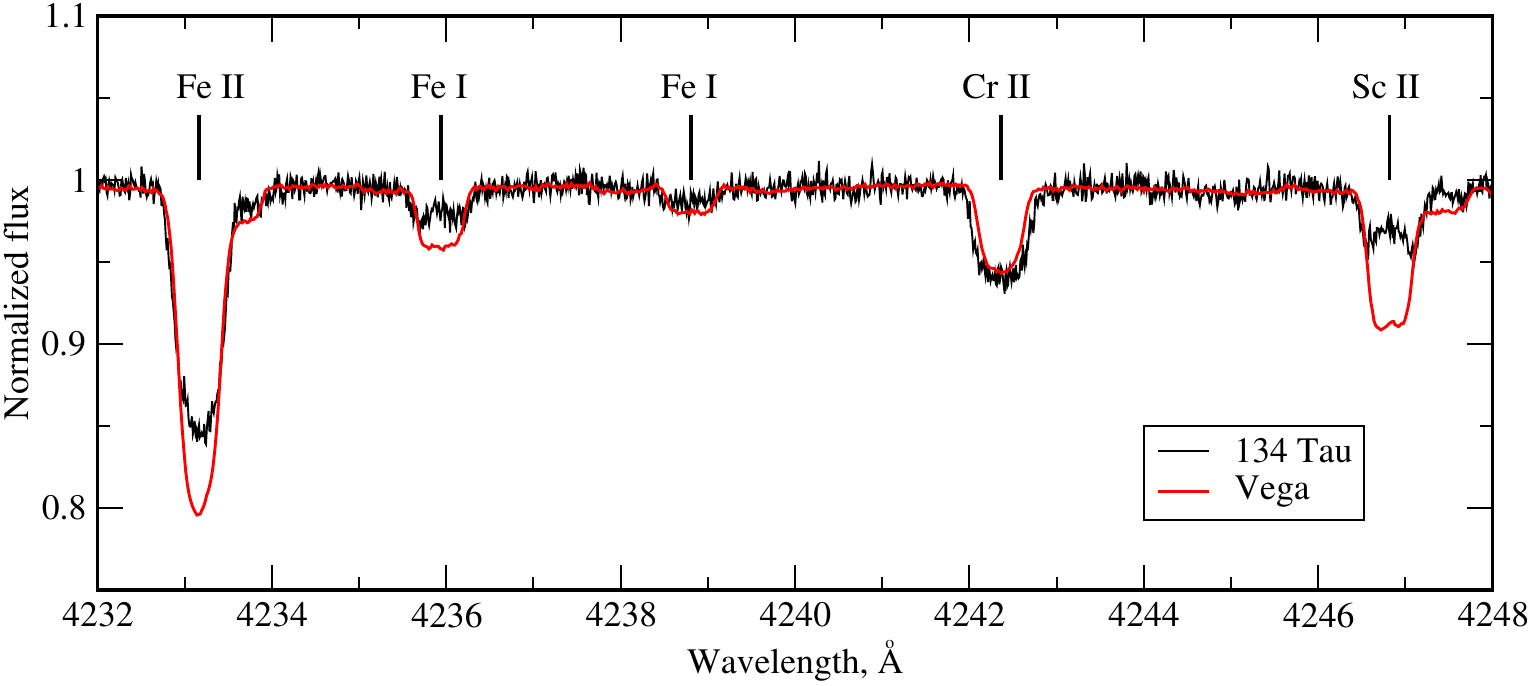}
     \caption{Comparison of the observed spectra of 134 Tau (black line) and Vega (red line). }
         \label{134Tau-Vega}
     \end{figure}

We estimated \teff\ = 12000~K as an upper limit for detection of the resonance Ba\ii\ 4554~\AA\ line in a spectrum of a star with \vsini\,= 25~\kms\ and Ba excess of +1~dex. The task becomes more difficult taking into account the fact that the mean projected rotational velocity of A and B-type stars exceeds 50 \kms. There is at least one star 134~Tau with \vsini\,= 27~\kms\ and \teff\ = 10825~K \citep{1991MNRAS.252..116A}, which was classified as a normal solar abundance star. We have extracted a spectrum of 134~Tau from the HARPS spectrograph archive\footnote{http://archive.eso.org/wdb/wdb/eso/repro/form}, examined it and found that spectral lines shape provides a strong evidence for 134~Tau being a fast rotator seen pole-on similar to Vega. This is demonstrated in Fig.~\ref{134Tau-Vega} where the spectrum of 134~Tau is compared with the Vega spectrum. More sophisticated methods are required for spectrum and abundance analysis that is beyond the scope of our paper. However, we estimated Sr and Ba abundances in 134~Tau based on the equivalent widths of Sr\ii\ 4077, 4215~\AA\ and Ba\ii\ 4554~\AA\ lines and using stellar parameters from \citep{1991MNRAS.252..116A}. Sr and Ba abundances are plotted in Fig.~\ref{Normal-Teff} by the open symbols. These results support the decreasing trend for the Sr and Ba excess in normal stars with \teff\ $>$ 10000~K.

\subsection{Am stars}
A comparison of the abundance patterns of \Peg\, and \Vir\, and known Am stars Sirius and HD~72660 is shown in Fig.~\ref{fig:op-tv-vs-sirius}. \Peg\, and \Vir\, show abundance patterns partially similar to hot representatives of mild Am stars: close to solar/small underabundance of light elements, and then a gradual increase of up to 1.5~dex in [X/H] towards Nd. Our sample of Am stars including the comparison stars has a narrow effective temperature range around 9700$\pm$150~K, and this results in close abundances for most elements. However, for the number of elements (C, N, K, Ca, Sc) the large abundance dispersion exceeding 0.5~dex is found, and it cannot be explained by the temperature effect.

CNO elements are moderately underabundant, but while O deficiency is practically the same, $\approx -0.25$~dex in our Am stars, C and N abundance dispersion reaches 0.7~dex.
Potassium abundance is measured in three stars of our sample, and it spreads over $\pm$0.3~dex around the solar value.
For Sirius, we estimated an upper limit for the potassium abundance by analysing an extremely weak K\ione\ 7698.96~\AA\ line. The estimated abundance was found to be close to the solar value and comparable to that observed in \Peg.

While the classical Am-star classification criteria (Ca and, to a greater extent, Sc abundance deficiency) are evident in Sirius, they are nearly absent in \Peg\ and \Vir. The expected positive NLTE abundance corrections for Sc may remove completely small Sc deficiency in \Peg. The NLTE analysis of Sc\ii\ lines in A type stars is urgently needed because Sc deficiency is considered as one of the classification characteristics of Am-stars. The difference between Ca and Sc abundances in Sirius and those observed in \Peg\ and \Vir\ cannot be explained by a temperature effect because three stars have close temperatures.
Obviously, classification characteristics of Am stars require more careful analysis based on detailed NLTE abundance analysis of a large sample of stars with small/moderate rotational velocities.

Nothing can be said about the temperature dependence of the neutron capture elements in our sample of Am stars because all four objects studied have similar effective temperatures. We may only conclude that both in normal and Am stars Ba has the largest abundance excess compared to nearby neutron capture elements such as Sr-Y-Zr and Nd.

\section{Evolutionary status}\label{evol}
\begin{figure}
         \centering
         \includegraphics[width=0.49\textwidth]{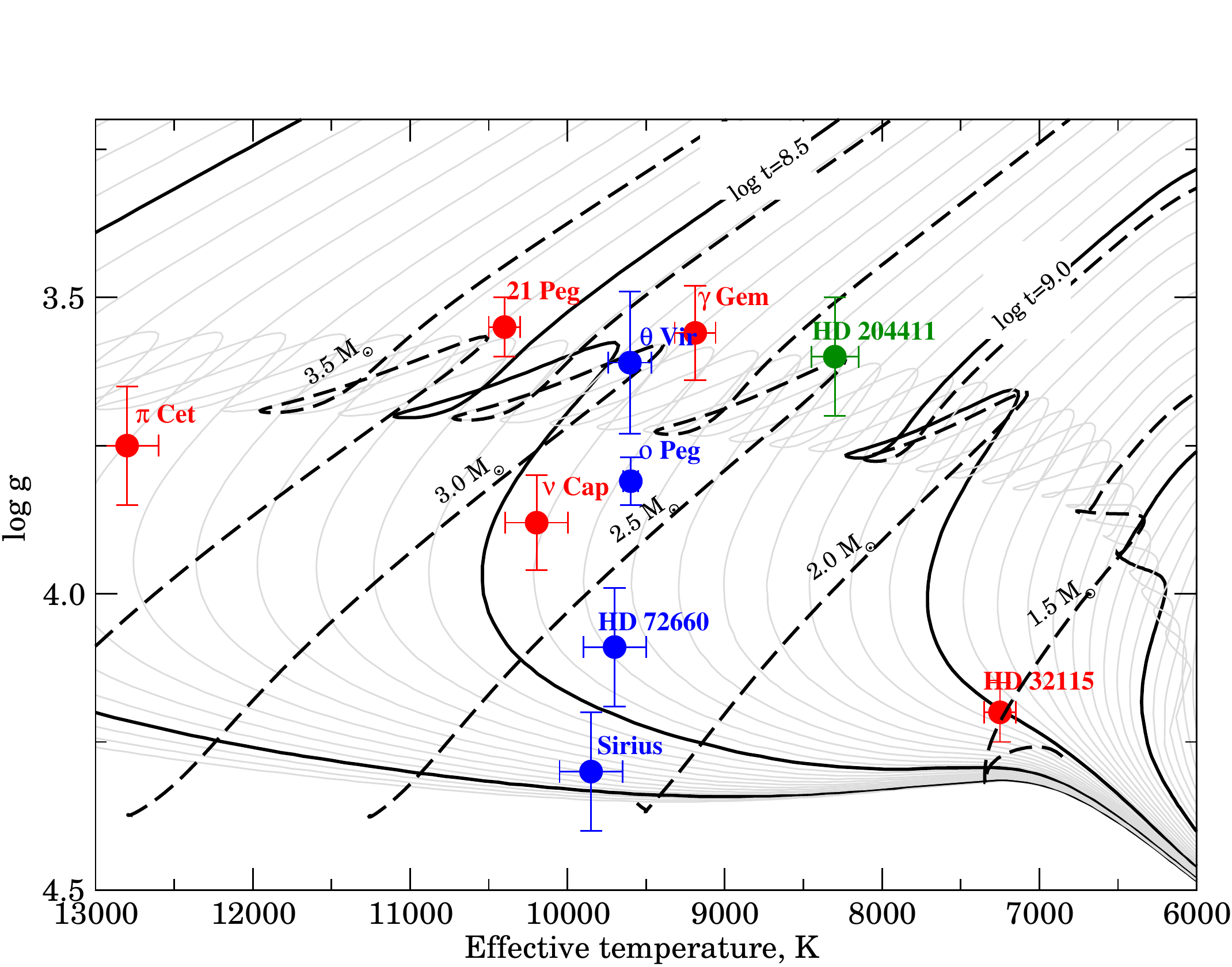}
       \caption{Position of the normal stars (red filled circles) and Am stars (blue filled circles) on the MESA isochrones (solid thick and thin lines) calculated with the solar metallicity. Evolutionary tracks are shown by dashed lines. Green filled circle indicates the position of the Ap star HD~204411.}
         \label{fig:iso}
\end{figure}

We estimated the evolutionary status of our stars putting their \lgg\ and \teff\ on the isochrones extracted from the single-star MESA \citep[Modules for Experiments in Stellar Astrophysics,][]{2016ApJ...823..102C}
theoretical stellar evolution model grid\footnote{\url{http://waps.cfa.harvard.edu/MIST/}}  computed with the solar-scaled abundance ratios (see Fig.~\ref{fig:iso}).  MESA has two variants for accounting for stellar rotation: V(initial)/V(critical)=0 and V(initial)/V(critical)=0.4. Our sample stars are slowly rotating objects so we use the first value. While, our stars has different abundance patterns, their average metallicities (z-values) do not differ significantly from the solar value. Therefore, we adopted MESA isochrones calculated with the solar metallicity. Three Am stars that have similar abundance patterns are sitting on the same isochrone with  an age of $\log{t}$ = 8.6 ($t\approx$400~Myr), while another Am star Sirius is younger, $t\approx$230~Myr.

The best way to explore evolutionary effects in creation of abundance anomalies is to study groups of slow rotating A-B stars in open clusters of different ages. According to the results by \citet{2022AJ....164..255C} who compared averaged abundance patterns of Am stars in four open clusters in age interval 100--700~Myr, abundances of 11 elements from O to Ba are consistent from cluster to cluster. 

\section{Comparison of A, Am, and Ap-type stars}\label{Comp2}
As we mention in the Introduction, the origin of the chemical anomalies is attributed to the atomic diffusion processes. These processes are slow, therefore a relatively stable atmosphere is necessary to support the element separation. Rapid rotation as well as turbulent motions will destroy any element separation homogenising atmospheric abundances. In principle, two slowly rotating stars with similar ages and atmospheric parameters should demonstrate similar atmospheric abundance distribution. However, it is well-known in the literature that there are A-type stars of various types exhibiting diverse abundance anomalies. The existence of different types of A stars suggests that diffusion is not the only mechanism responsible for abundance peculiarities. At least, in magnetic chemically peculiar (Ap) stars, diffusion is considered to be the primary cause of abundance anomalies.

To estimate the contribution of diffusion to abundance patterns of A and Am sample stars, we included for comparison Ap star HD~204411 with \teff\ = 8300~K, \lgg\ = 3.6 \citep[][]{2019MNRAS.488.2343R}, and a weak averaged surface magnetic field modulus \bs$\le$0.8~kG, which has only a minor effect on abundances. Similar to \Gem\, HD~204411 is approaching the end of its main sequence life (Fig.~\ref{fig:iso}). 
\Gem\, and \Vir\, are selected as representatives of A and Am stars, respectively. Additionally, HD~72660 is included as an Am star with a different mass. 

\begin{figure}
         \centering
         \includegraphics[width=0.49\textwidth]{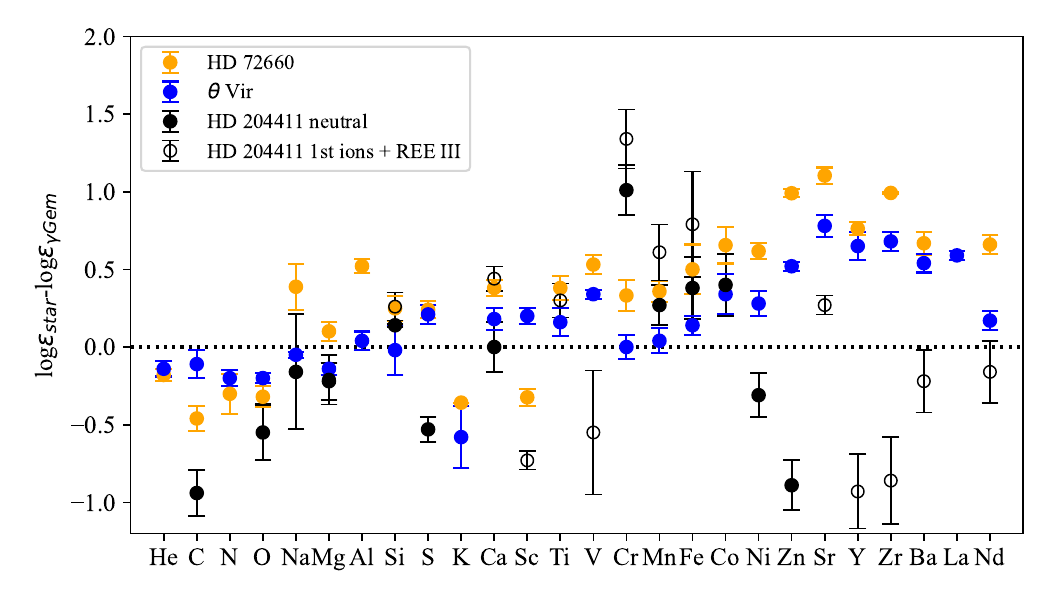}
     \caption{Differential abundances in Am stars HD~72660, \Vir\ and Ap star HD~204411 relative to \Gem.}
         \label{fig:Astars-vs-GG}
\end{figure}

For comparison of the abundance pattern in different stars we adopt \Gem\ as the reference star. To achieve more precise matching, we calculate line-by-line differential abundances with respect to \Gem\ for the  selected stars. For Ap star HD~204411, we took LTE abundances from \citet{2019MNRAS.488.2343R} and applied the NLTE abundance corrections for C, O, Na, Mg, Si, Ca, Zn, Sr, Y, Zr, Ba. Fig.~\ref{fig:Astars-vs-GG} shows abundance difference between the Am stars \Vir\ and HD~72660, and Ap star HD~204411 relative to the normal star \Gem. 

In Ap star HD~204411, atomic diffusion produces the observed stratification of Ca, Cr, Fe \citep{2019MNRAS.488.2343R}, and it also appeared as the violation of the ionisation balance (Fig.~\ref{fig:Astars-vs-GG}): there is a significant difference in abundances from the lines of neutral species and the first ions of the same chemical element.  In contrast to HD~204411, the ionisation balance is fulfilled in our A and Am stars.

All neutron capture elements except Sr have lower abundances in Ap star HD~204411 compared to the normal A star, and it is Sr, not Ba that has the largest abundance among elements Zn-Sr-Y-Zr-Ba. It should be noted that in Ap stars with strong magnetic fields of ~4 kG, an  excess of neutron capture elements of  about 2~dex is observed, with barium showing the lowest abundance among all neighbouring elements  \citep[see Fig.~1.6a in][]{Romanovskaya_2022}. 

Abundances of Fe-peak element, excluding Cr, in Ap star are close to that in Am star HD~72660. Cr is the most abundant among Fe-peak elements and its overabundance together with Sr is one of the classification criterium of Ap stars (SrCr stars). 

But what defines a difference between normal A and Am star? It is possible that in Am stars the chemical composition is affected by the presence of a secondary component: Sirius has a white dwarf companion;  HD~72660 is a suspected single-line spectroscopic binary \citep{2014A&A...562A..84R}, \Peg\, and \Vir\, are double or multiple stars (see Section~\ref{sec1:star} for details). The binarity of HD~72660 was confirmed by the observed proper motion anomaly \citep{2019A&A...623A..72K}. However, a normal star \Gem\ is also a spectroscopic binary. Although the fast rotation should prevent atomic diffusion process, however if 134~Tau is confirmed to be normal pole-on fast rotating A star, then Sr and Ba overabundances is difficult to explain by this mechanism.

During the comparison analysis, the following questions arise:\\
$\bullet$\ what distinguishes normal A type stars from Am stars? \\ 
$\bullet$\ which mechanisms are responsible for abundance diversity among A type stars with close fundamental parameters?\\  
$\bullet$\ do all A stars have non-solar abundances of neutron capture elements?\\
$\bullet$\ are there normal A type stars with solar abundances? \\

The first two questions can be addressed by theoretical studies  of stellar evolution and modelling of processes in A-type stars, which lie beyond the scope of this study. Here we aim to provide as accurate as possible observational constraints using a careful stellar parameter determination and state of the art spectroscopic analysis.  We present our comments regarding the possible impact of a secondary component on Am-type stars phenomenon. 

The latter questions are attributed to our findings concerning the excess of heavy elements and its correlation with the effective temperature. In this study, we confirm the hypothesis proposed in Paper \ione, which suggests that the excess of heavy elements in A-type stars increases with \teff. In addition to these conclusion, here we found that the heavy element excess reaches its maximum at \teff\ = 10000~K,  and as \teff\ increases further, the element excess gradually decreases until it reaches the solar value at \teff\ = 13000 K. This indicates that stars with solar element abundances can be found among late B-type stars with \teff\ > 13000 K.

\section{Conclusions}
\label{conclusions}
We present a self-consistent abundance analysis for elements from He to Nd in slowly rotating normal A and metallic line Am stars in temperature range 9200--10200 K. We refined their fundamental parameters with the SME package, spectral energy distribution, and the hydrogen line profiles. We analysed the chemical abundances of 26 elements, for 18 of them the analysis was performed with NLTE approach. 

Our abundance analysis supports the classification of \Gem\ and \Cap\ as normal A stars with enhanced neutron capture element abundances and \Peg\ and \Vir\ as Am stars. We found that, in NLTE, the abundance difference between stars of the same type reduces.

For normal A-type sample stars, abundances of elements from He to Ni are close to solar values, while elements heavier than Zn are overabundant with Ba showing the largest excess. We confirm a gradual rise of the heavy element (Zn, Sr, Y, Zr, Ba) abundance excess up to +1 dex with \teff\ increasing from 7200 to 10000~K noted first in Paper \ione. We found that with a further increase in \teff, a decreasing trend can be noticed, and the Sr excess mostly reaches zero (solar abundance) at \teff\ = 12800~K. To support our present result, abundance studies of more stars with slow rotation and solar metallicity in temperature ranges 8000--9000~K and 9500--12000~K are required.

For metallic line Am stars of our sample, abundance patterns are similar to hot representatives of mild Am stars: close to solar/small underabundance of light elements, and then a gradual increase of up to 1.5~dex in [X/H] towards Nd. Our sample stars do not show Ca-Sc deficiency that is pronounced in Sirius. In contrast, a small Ca excess is observed in \Peg, \Vir, and HD~72660. The expected positive NLTE abundance corrections for Sc may remove completely a small Sc deficiency in \Peg, and, perhaps, in HD~72660. Therefore, the classification criteria of Am stars based on significant Ca and Sc deficiency cannot be applied. A comprehensive NLTE abundance analysis of a substantial stellar sample is necessary to reevaluate the classification criteria of Am stars.

Comparison of the atmospheric abundances in the representatives of different types of A stars (normal, Am and Ap) with similar fundamental parameters showed a diversity in the abundance patterns that cannot be explained by a simple atomic diffusion process.

\section*{Acknowledgements}

We are indebted to  L.~I. Mashonkina for the NLTE calculations and for useful comments on this study. We thank D.~V. Shulyak for the help in \textsc{LLmodels} grids calculations, and the anonymous referee for very fruitful comments improving the present paper.
This work has made use of data from the European Space Agency mission Gaia (https://www.cosmos.esa.int/gaia), processed by the Gaia Data Processing and Analysis Consortium (DPAC, https://www.cosmos.esa.int/web/gaia/dpac/consortium).

\section*{Data Availability}
\label{dataavailability}
 
Spectra of program stars normalised to the continuum level are provided as supplementary material in the form of an ASCII file in CDS. The full table of used lines with log\,$gf$,  excitation potentials of lower level $E_i$, LTE and non-LTE abundances and references for oscillator strengths and hyperfine constants is available online in CDS.



\bibliographystyle{mnras}
\bibliography{Ref_Romanovskaya_2022,ts_a2.bib} 



\appendix

\section{Spectral energy distributions figures}
This section presents the spectral energy distributions for the stars \Peg, \Vir, \Cap. See details in section \ref{model}.
\vspace{1cm}

\begin{figure*}
    \centering
    \includegraphics[width=0.95\textwidth, clip]{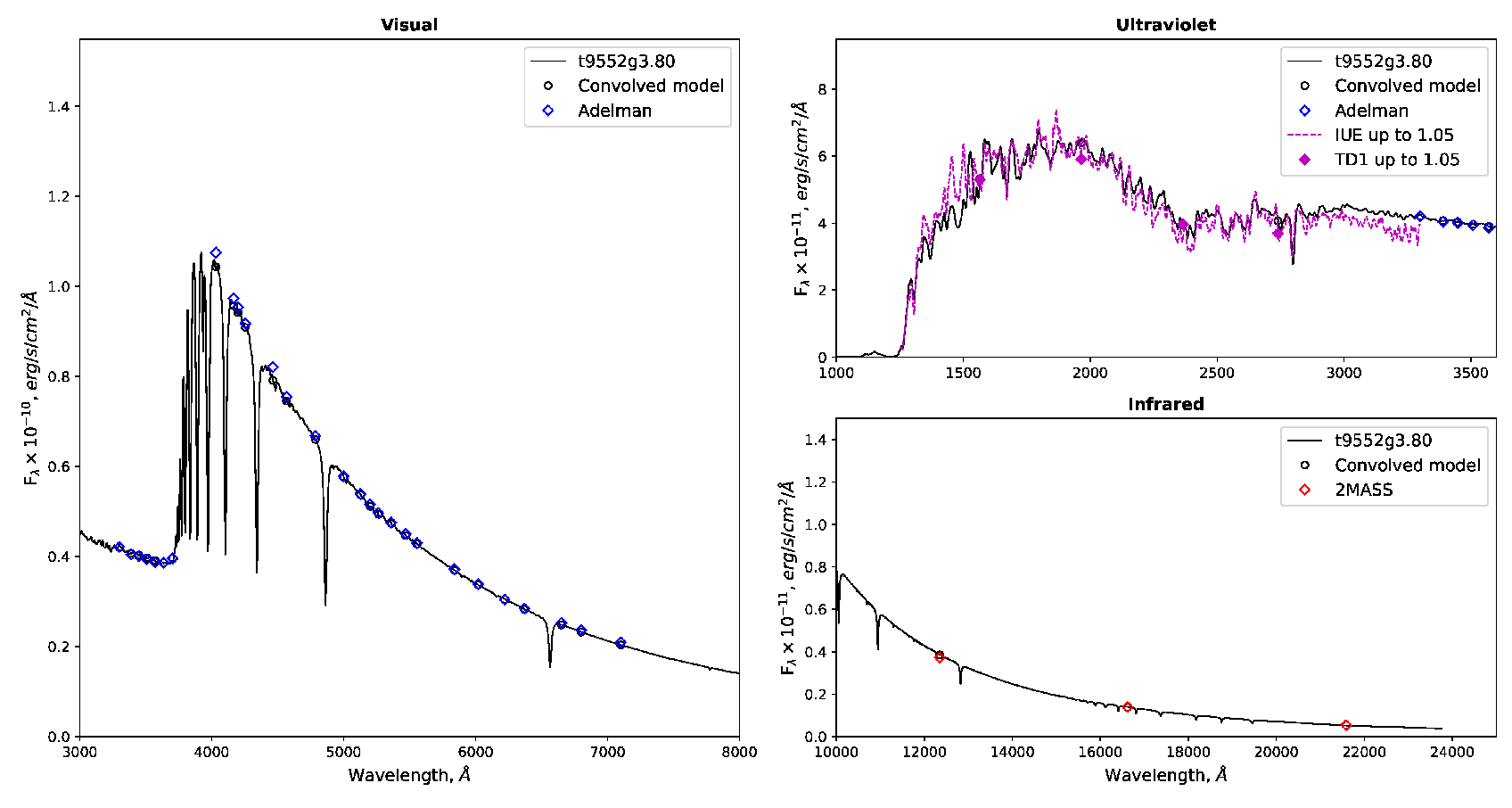}
    \caption{Spectral energy distribution of \Peg\, in the UV, visual and IR spectral regions. Black solid line shows theoretical \SED\, calculation with the model atmosphere parameters t9552g38. IUE and TD1 points were raised by a factor of 1.05.}
    \label{fig:omipeg-sed}
\end{figure*}

\begin{figure*}
    \centering
    \includegraphics[width=0.95\textwidth, clip]{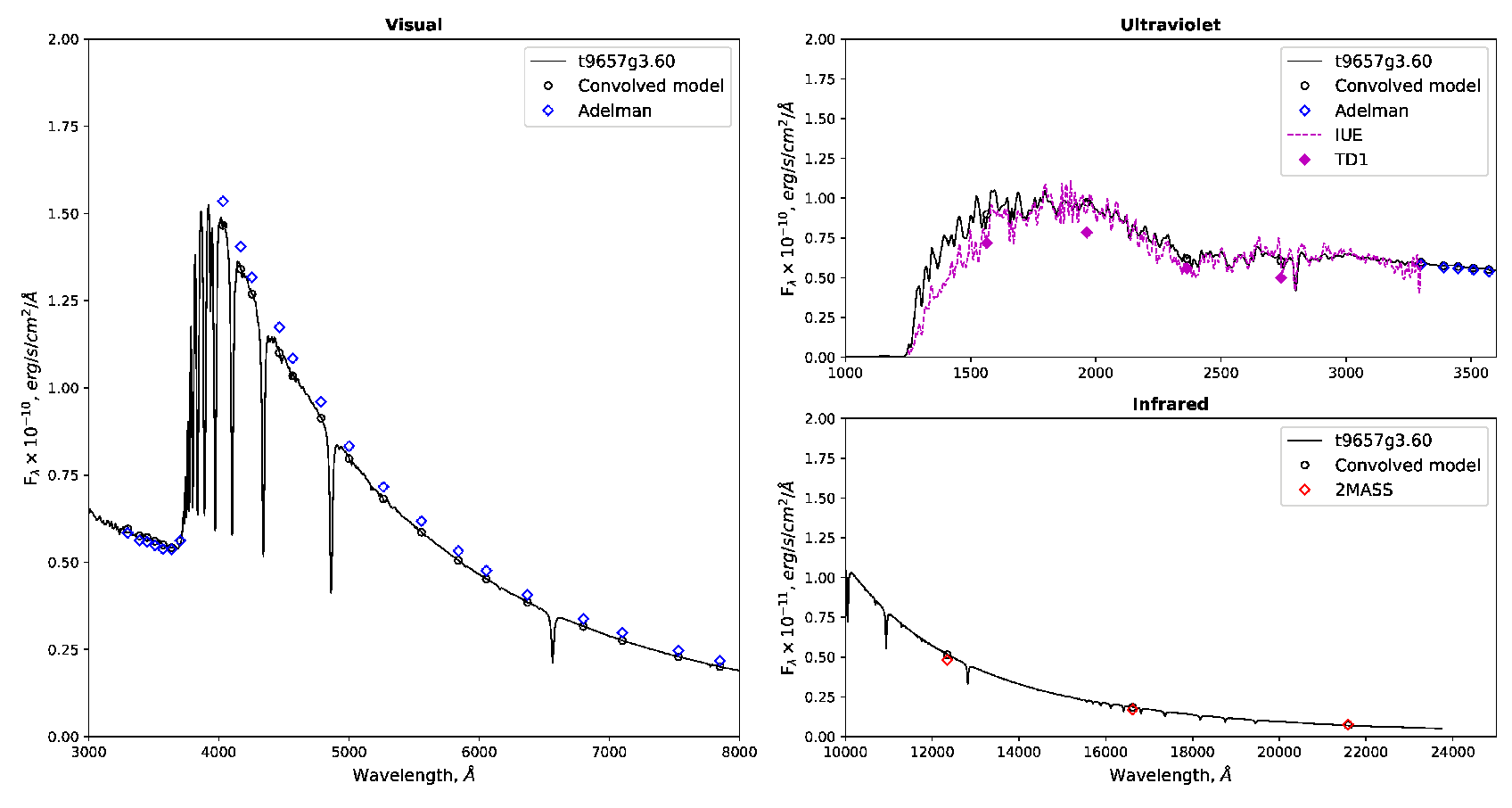}
    \caption{Spectral energy distribution of \Vir\, in the UV, visual and IR spectral regions. Black solid line shows theoretical \SED\, calculation with the model atmosphere parameters t9657g36.}
    \label{fig:thetavir-sed}
\end{figure*}

\begin{figure*}
    \centering
    \includegraphics[width=0.95\textwidth, clip]{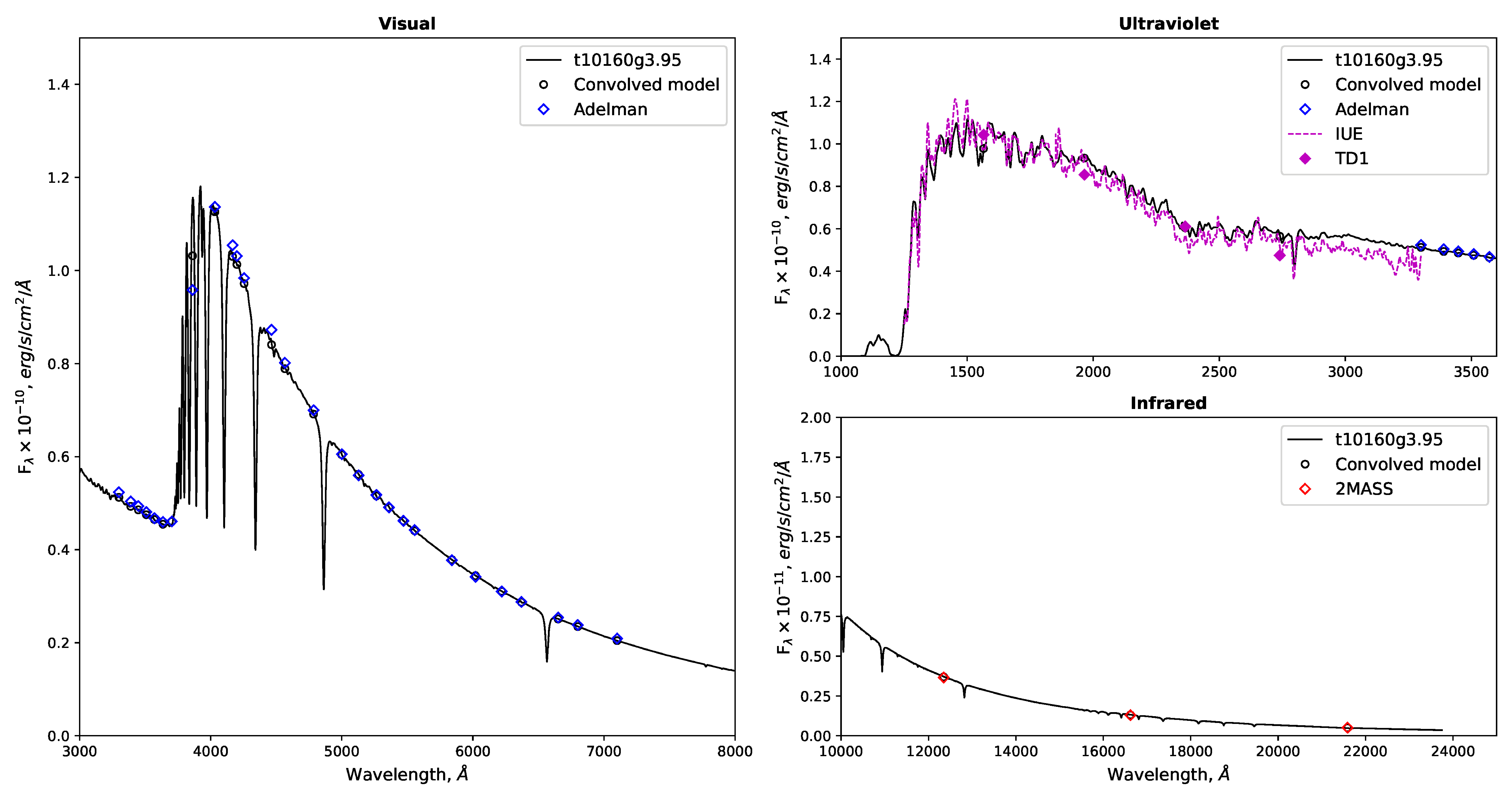}
    \caption{Spectral energy distribution of \Cap\, in the UV, visual and IR spectral regions. Black solid line shows theoretical \SED\, calculation with the model atmosphere 
				parameters t10160g395.}
    \label{fig:nucap-sed}
\end{figure*}

\bsp	
\label{lastpage}
\end{document}